\documentclass[11pt,a4paper,fleqn,usenatbib]{mnras}

\usepackage{newtxtext,newtxmath}

\usepackage{ulem}

\usepackage[T1]{fontenc}
\usepackage{ae,aecompl}

\usepackage{graphicx}	
\usepackage{amsmath}	
\usepackage{amssymb}	

\newcommand{\Msun}{\, {\rm M_{\odot}}}

\title[Self-interacting dark matter and Milky Way dwarf galaxies]{Constraining Velocity-dependent Self-Interacting Dark Matter with the Milky Way's dwarf spheroidal galaxies}
\author[Camila A. Correa]{
Camila A. Correa$^{1}$\thanks{E-mail: camila.correa@uva.nl}
\\
$^{1}$GRAPPA Institute, University of Amsterdam, Science Park 904, 1098 XH Amsterdam, The Netherlands
}

\date{Accepted XXX. Received YYY; in original form ZZZ}

\pubyear{2015}

\begin{document}
\label{firstpage}
\pagerange{\pageref{firstpage}--\pageref{lastpage}}
\maketitle

\begin{abstract}
The observed anti-correlation between the central dark matter (DM) densities of the bright Milky Way (MW) dwarf spheroidal galaxies (dSphs) and their orbital pericenter distances poses a potential signature of self-interacting dark matter (SIDM). In this work we investigate this possibility by analysing the range of SIDM scattering cross section per unit mass, $\sigma/m_{\chi}$, able to explain such anti-correlation. We simulate the orbital evolution of dSphs subhaloes around the MW assuming an analytical form for the gravitational potential, adopting the proper motions from the Gaia mission and including a consistent characterization of gravitational tidal stripping. The evolution of the subhaloes density profile is modelled using the gravothermal fluid formalism, where DM particle collisions induce thermal conduction that depends on $\sigma/m_{\chi}$. We find that models of dSphs, such as Carina and Fornax, reproduce the observed central DM densities with fixed $\sigma/m_{\chi}$ ranging between $30$ and $50$ cm$^{2}$g$^{-1}$, whereas other dSphs prefer larger values ranging between $70$ and $100$ cm$^{2}$g$^{-1}$. These cross sections correlate with the average collision velocity of DM particles within each subhalo's core, so that systems modelled with large cross sections have lower collision velocities. We fit the cross section-velocity correlation with a SIDM particle model, where a DM particle of mass $m_{\chi}=0.648\pm 0.154$ GeV interacts under the exchange of a light mediator of mass $m_{\phi}=0.636\pm 0.055$ MeV, with the self-interactions being described by a Yukawa potential. The outcome is a cross section-velocity relation that explains the diverse DM profiles of MW dSph satellites and is consistent with observational constraints on larger scales.
\end{abstract}



\section{Introduction}

The standard cosmological paradigm of $\Lambda$ collisionless cold dark matter ($\Lambda$CDM) accurately predicts the large-scale structure of the Universe (\citealt{Springel06}), however its success is less certain over the small scales, the regime relevant to the substructure within galactic haloes. The deficit of observed low-mass satellite galaxies within the Local Group relative to predictions from analytical theory (\citealt{Press74}) and CDM N-body simulations (\citealt{Klypin99}), along with the discrepancy between the low dark matter densities of some galaxies and the cuspy and dense subhaloes predicted by simulations (e.g. \citealt{Flores94,Moore94,Moore99}), motivated to question the cold and collisionless nature of dark matter (DM). The alternative was that DM might have non-negligible self-interactions (\citealt{Spergel00}).

Self-interacting dark matter (hereafter SIDM) assumes that DM particles experience collisions with each other, these collisions transfer heat towards the colder central regions of DM haloes, lowering central densities and creating constant density cores (e.g. \citealt{Dave01,Colin02,Vogelsberger12,Rocha13,Dooley16,Vogelsberger19,Robles19}). DM particle collisions also lead to less concentrated subhaloes that are more prone to tidal disruption, as well as to the evaporation of subhaloes via ram pressure stripping exerted by the larger host. In this manner various SIDM models have produced halo mass functions that are significantly suppressed on small scales (e.g. \citealt{Buckley14,CyrRacine16,Vogelsberger16}, who also assumed a suppression in the power spectrum, and e.g. \citealt{Nadler20} that assumed standard SIDM).

An important disagreement between the prediction of pure CDM simulations and observations is the so-called too-big-to-fail problem (hereafter TBTF), which states that the most massive subhaloes in CDM simulations are too dense in the centre to host the observed satellites of the Milky Way (\citealt{BoylanKolchin11,BoylanKolchin12}). The TBTF problem, although solved by invoking baryonic physics and environmental effects (e.g. \citealt{Sawala16,Dutton16,Wetzel16,Fattahi16}), has been revisited in the SIDM paradigm due to its prevalence in other galaxies besides the Milky Way (e.g. \citealt{GarrisonKimmel14,Papastergis15}). 

The SIDM models that alleviate the TBTF problem require a DM interaction cross section per unit mass, $\sigma/m_{\chi}$, to be larger than $\sigma/m_{\chi}{>}1$ cm$^{2}$g$^{-1}$ (\citealt{Zavala13}). Despite its success in solving the TBTF problem and other small-scales discrepancies (e.g. \citealt{Rocha13,Kamada17,Ren19}), the excitement caused by SIDM diminished due to the strong constraints set by X-ray and lensing observations of galaxy clusters, that set an upper-limit on the scattering cross section of the order of 1 cm$^{2}$g$^{-1}$ (\citealt{MiraldaEscude02,Peter13}). This upper-limit was later supported by observations of bounds from major mergers (\citealt{Randall08,Harvey15,Wittman18}) and bright central galaxy wobbles (\citealt{Kim17,Harvey19}).

An exciting alternative for SIDM, however, is that the cross section may depend on the relative velocity of DM particles, in such a way that DM behaves as a collisional fluid on small scales while it is essentially collisionless over large scales. Such velocity-dependence is supported by many theoretical models that argue that DM exists in a `hidden sector', where forces between DM particles are mediated by analogues to electroweak or strong forces (e.g. \citealt{Pospelov08,ArkaniHamed09,Buckley10,Feng10,Boddy14,Tulin18}). Velocity-dependent SIDM models have been explored on galaxy cluster scales (e.g., \citealt{Robertson17,Robertson19,Banerjee20}) and Milky Way (MW)-mass systems (e.g. \citealt{Vogelsberger12,Zavala13,Nadler20}). 

The works of \citet{Read18} and \citet{Valli18} analysed the density profiles of the MW dwarf spheroidal galaxies (dSphs), aiming to place constraints on $\sigma/m_{\chi}$ on dwarf galaxy scales. \citet{Read18} focused on Draco and claimed that its high central density gives an upper bound on the SIDM cross section of $\sigma/m_{\chi}{<}0.57$ cm$^{2}$g$^{-1}$. \citet{Valli18} used a similar methodology as \citet{Read18}, although they also found that Draco's high central density could be described by a $\sigma/m_{\chi}{\sim}0.4$ cm$^{2}$g$^{-1}$, they concluded that the remaining dSphs probed different cross sections, ranging between 0.1 and 40 cm$^{2}$g$^{-1}$. 

Recently, \citet{Kaplinghat19} have revisited the TBTF problem of MW dSphs. They have reported an interesting anti-correlation between the central DM densities of the bright dSphs and their orbital pericenter distances, so that the dSphs that have come closer to the MW centre are more dense in DM than those that have not come so close. \citet{Read19} proposes that the anti-correlation is the result of baryonic effects. The gas expelled by stellar feedback `heats' the surrounding DM, lowering the haloes' central density (\citealt{Navarro96}). If the effect repeats over several cycles of star formation, it accumulates, leading eventually to transform the DM cusp into a core (e.g. \citealt{Read05,Pontzen12}). However, for dwarf galaxies, such as Draco and Ursa Minor, that are DM-dominated and stopped forming stars long ago ($\sim 10$ Gyr), the DM cusp is formed again, thus explaining the high DM central densities. \citet{Read19} modelled the stellar kinematics to infer the DM distribution of the MW dSphs, and showed that all dwarfs except for Fornax, are well fitted by a cuspy profile. Even if the `lack' of baryonic effects is responsible for the high DM densities in the dSphs, it does not explain the origin of the anti-correlation with pericenter distance. 

SIDM, on the contrary, can potentially explain the observed anti-correlation. DM collisions lead to an outward heat transfer that induces gravothermal core collapse, i.e. the central density increases with time (\citealt{Balberg02,Elbert15}). This gravothermal collapse would be accelerated by mass loss via tidal stripping (\citealt{Nishikawa19}) and correlate with how close the satellite galaxies come to the centre of the MW. According to this scenario, the anti-correlation could not only be a potential signature of SIDM (\citealt{Kaplinghat19,Nishikawa19, Sameie20}), but it would also invalidate the upper limits on $\sigma/m_{\chi}$ from \citet{Read18} and \citet{Valli18}, since their analysis does not include the effects of gravothermal collapse.

The goals of this study are to investigate the possibility that the anti-correlation is a signature of SIDM, and analyse the ranges of $\sigma/m_{\chi}$ that could potentially explain it. To do so, we simulate the orbital evolution of dSphs subhaloes around the MW by adopting the proper motions from the Gaia mission (\citealt{Helmi18,Fritz18}), assuming an analytical form for the MW gravitational potential and including a consistent characterization of gravitational tidal stripping. The evolution of the density profile of SIDM subhaloes is simulated using the gravothermal fluid formalism, which was originally developed to study the evolution of globular clusters (\citealt{LyndenBell80}), but it has also been applied to isolated SIDM haloes (\citealt{Balberg02,Koda11,Shapiro18}). This method allows us to track the DM halo evolution within scales smaller than 100 pc, which are largely expensive to resolve with N-body simulations, as well as to easily cover a wide range of parameter space.

This work is organised as follows. Section \ref{SIDM_model} outlines our model setup. We present our results in Section \ref{Results} and compare to observational data. We discuss constraints on the cross section-velocity relation in Section \ref{Sec_Velocity_cross_section}. Comparison with previous works, as well as the challenges of the model and impact of initial conditions are discussed in Section \ref{Sec_Discussion}. Finally, Section \ref{Conclusions} summarises our key results.



\section{SIDM halo model}\label{SIDM_model}

The SIDM halo model derived in this work connects the gravothermal fluid approximation, with orbit integration and tidal stripping modelling. Subsection~\ref{Sec_Gravothermal_collapse} describes the gravothermal fluid formalism, the orbital evolution is introduced in Subsection~\ref{Sec_Orbital_evolution} and Subsection~\ref{Sec_Tidal_Stripping} describes the gravitational tidal stripping modelling. We summarise the complete model and describe the initial conditions in Subsection~\ref{Sec_integration_of_eqs}.

\subsection{Gravothermal collapse}\label{Sec_Gravothermal_collapse}

We consider a spherical halo in isolation and in quasi-static virial equilibrium, with a density profile $\rho(r,t)$ and an enclosed mass of $m({<}r,t)$ at radius $r$ and time $t$. We assume that the ensemble of gravitating particles is well approximated by a fluid-like description, where the effective temperature is identified with the square of the one-dimensional velocity dispersion, $v(r,t)$, and thermal heat conduction is employed to reflect the manner in which the close-encounter large-angle (hard-sphere) scatterings combine to transfer energy in the system. The quasi-static approximation means that, while the fluid evolves thermally, it always satisfies hydrostatic equilibrium at each moment. 

The fundamental equations of the model are mass conservation, hydrostatic equilibrium, energy flux equation, and the first law of thermodynamics, 

\begin{eqnarray}\label{mass_conservation_eq}
\frac{\partial m}{\partial r} &=& 4\pi r^{2}\rho,\\\label{hydro_equi_eq}
\frac{\partial (\rho v^{2})}{\partial r} &=& -\frac{Gm\rho}{r},\\\label{flux_eq}
\frac{L}{4\pi r^{2}}&=&-\kappa \frac{\partial T}{\partial r},\\\label{first_law_eq}
\frac{\partial L}{\partial r} &=& -4\pi r^{2}\rho v^{2}\left(\frac{\partial}{\partial t}\right)_{m}\log\left(\frac{v^{3}}{\rho}\right),
\end{eqnarray}

\noindent where $L(r)$ the luminosity through a sphere at $r$.

The flux equation (eq.~\ref{flux_eq}) can be written into a single expression that considers both the cases where the the mean free path between collisions is significantly shorter (known as SMFP regime) or larger (LMFP regime) than the system size (\citealt{Balberg02}), as follows 

\begin{equation}\label{flux_eq_2}
\frac{L}{4\pi r^{2}} = -\frac{3}{2}b\rho v\left[\left(\frac{1}{\lambda}\right)+\left(\frac{vt_{r}}{CH^{2}}\right)\right]^{-1}\frac{\partial v^{2}}{\partial r},
\end{equation}

\noindent where $H\equiv \sqrt{v^{2}/(4\pi G\rho)}$ is the gravitational scale height of the system, $\lambda$ is the collisional scale for the mean free path given by $\lambda=1/(\rho\sigma_{m})$, with $\sigma_{m}=\sigma/m_{\chi}$ the cross section per unit mass, and $t_{r}\equiv \lambda/(av)$ is the relaxation time, with $a=\sqrt{16/\pi}$ for hard-sphere scattering of particles with a Maxwell-Boltzmann velocity distribution (\citealt{Balberg02}). 

In the SMFP regime the continuum assumption applies and the collection of particles can accurately be treated as a continuous fluid. From this regime, the effective impact parameter $b=25\sqrt{\pi}/32\approx 1.38$ in eq. (\ref{flux_eq_2}) can be derived from first principles (e.g. \citealt{Chapman90}). In the LMFP regime the model needs to be calibrated using N-body simulations. Previous studies have tested the parameter $C$ that determines the radial heat conduction for isolated (\citealt{Balberg02,Koda11,Essig19}) and cosmological N-body simulations (\citealt{Elbert15,Nishikawa19,Essig19}) with purely elastic DM self-interactions. 

\citet{Essig19} and \citet{Nishikawa19} assumed spherical symmetry but not isolation. \citet{Essig19} used the cosmological simulation from \citet{Elbert15} and showed that $C=0.45-0.6$ closely reproduces the density and velocity dispersion from the simulation. \citet{Nishikawa19} also used \citet{Elbert15} simulation and concluded that for $\sigma/m_{\chi}>10$ cm$^{2}$g$^{-1}$, $C=0.75$ and $b=0.003$ is needed to reproduce the subhalo's DM density. Both studies also reported differences with respect to the simulation in the subhalo density of up to a factor of 2 for $\sigma/m_{\chi}=50$ cm$^{2}$g$^{-1}$. We take this discrepancy into account when we constrain the cross section in Section~\ref{Sec_Velocity_cross_section}, and adopt $C=0.75$ as reported by \citet{Balberg02}, who assumed spherical symmetry and isolation for the modelling of SIDM haloes. However, we also discuss how the different values of $C=0.45-0.6$ and $b=0.003-1.38$ impact on our key results in Subsection~\ref{Sec_Caveats}.

\subsection{Orbital evolution of MW spheroidal galaxies}\label{Sec_Orbital_evolution}

Throughout this work we model the internal dynamics and orbital evolution of the nine most luminous MW dSph galaxies. These include Ursa Minor (hereafter UM), Draco, Sculptor, Sextans, Fornax, Carina, LeoII and LeoI, and Canes Venatici I (hereafter CVnI). We focus on these systems because they have the highest quality kinematic data and the largest samples of spectroscopically confirmed member stars to resolve the dynamics at small radii. 

The second data released by the Gaia mission (\citealt{Brown18,Helmi18}) has largely increased the precision and amount of astrometric data of Galactic stars, making possible the determination of spatial motions of many dSphs orbiting the MW halo (\citealt{Helmi18,Fritz18}). Using the proper motions determined by \citet{Fritz18}, and the publicly available code galpy\footnote{http://github.com/jobovy/galpy} (\citealt{Bovy15}), we integrate the orbits of the dSphs adopting the static MWPotential14 model, which has been shown to be consistent with various observations (see \citealt{Bovy15} for details). For the MW dark matter halo mass we assume a virial mass of $M_{200}=10^{12}\Msun$, defined as the total within $R_{200}$, radius within which the mean density is equal to 200 times the critical density of the Universe, $\rho_{\rm{crit}}$. In Appendix \ref{Comparison_MWMass_appendix} we show that assuming a lighter MW halo mass of $0.8\times 10^{12}\Msun$ or a heavier model of $1.6\times 10^{12}\Msun$ does not largely affect our key results. 

\begin{figure} 
	\includegraphics[angle=0,width=0.48\textwidth]{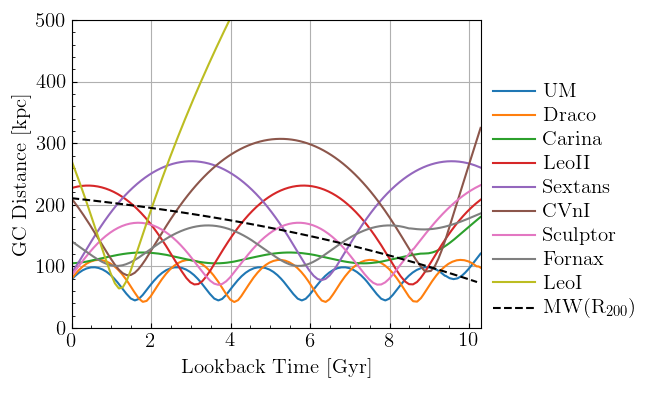}
	\vspace{-0.5cm}
	\caption{Time evolution of the galactocentric distance of the MW dSphs over the past 10 Gyrs. The black dashed line indicates the time evolution of the virial radius of the MW's dark matter halo, $R_{200}$, which increases with time under the growing mass of the dark halo. Note, however, that the integration of the dSphs orbit was done assuming the MWPotential14 static model from Body et al. (2015).}
	\label{OrbitsDwarfGalaxies}
\end{figure}

Fig.~\ref{OrbitsDwarfGalaxies} shows the time evolution of the galactocentric distance of the MW dSph galaxies. The black dashed line indicates the time evolution of the MW's virial radius, $R_{200}$, calculated using the halo accretion history model

\begin{equation}\label{MAH_model_Correa}
M(z)=M(z=0)(1+z)^{\alpha}e^{\beta z},
\end{equation}

\noindent from \citet{Correa15a,Correa15b}\footnote{https://camilacorrea.com/code/commah/}. It can be seen from Fig.~\ref{OrbitsDwarfGalaxies} that the dSphs of UM, Draco, Sculptor and Carina became satellites of the MW nearly 8-9 Gyrs ago and since then they have completed many orbits, whereas LeoII, Sextans, CVnI and Fornax have completed two orbits around the MW. LeoI crossed the MW's virial radius for the first time roughly 2 Gyrs ago.

\subsection{Gravitational tidal stripping}\label{Sec_Tidal_Stripping}

An important aspect in the evolution of small subhaloes relative to haloes in the field, is that when subhaloes are accreted by a larger halo, they begin to lose mass due to the strong gravitational tidal interactions exerted by the larger host. Studies using numerical simulations have shown that gravitational tidal stripping can even lead to the complete disruption of a large fraction of subhaloes (\citealt{Han16,Jiang17,vandenBosch17,vandenBosch18}), likely enhanced in the presence of SIDM (\citealt{Dooley16,Nadler20}). 

We calculate the rate of mass loss, $dm/dt$, of the subhaloes hosting the dSphs as they orbit around the MW by adopting the following tidal stripping rate

\begin{equation}\label{TidalStripping}
\frac{dm}{dt} = \frac{m({>}r_{t})}{\tau_{\rm{orb}}/\alpha},
\end{equation}

\noindent where $m({>}r_{t})$ is the subhalo mass outside the instantaneous tidal radius $r_{t}$, $\tau_{\rm{orb}} = 2\pi/\omega$ with $\omega$ the instantaneous angular velocity of the subhalo, and $\alpha=1$ (see \citealt{vandenBosch18}). The tidal radius is calculated as

\begin{equation}\label{TidalRadius}
\left(\frac{r_{t}}{R}\right)^{3} = \frac{m({<}r_{t})/M({<}R)}{2+\frac{\Omega^{2}R^{3}}{GM(R)}-\frac{d{\rm{ln}} M}{d{\rm{ln}} R}|_{R}},
\end{equation} 

\noindent which corresponds to the scenario where a subhalo of mass $m$ is on a circular orbit of radius $R$, with angular speed $\Omega(=V_{\rm{circ}}(R)/R)$, around a halo of mass $M$ (e.g., \citealt{King62,Tollet17}).

Eq.~(\ref{TidalStripping}) gives the approximate amount of mass stripped from the subhalo over a short time-step, but it does not indicate how the density profile is modified by it. To model the truncation in the density profile, we employ the fitting functions of \citet{Green19}, that follow the structural evolution of a tidally truncated subhalo and solely depend on the fraction of mass stripped.

\citet{Green19} used the DASH library (\citealt{Ogiya19}) of high-resolution, idealized dark matter only collisionless N-body simulations that follow the evolution of an individual subhalo as it orbits within the fixed, analytical potential of its host halo. Both the fixed host halo and the initial subhalo are spherically symmetric, each with a Navarro-Frenk-White (hereafter NFW, \citealt{NFW97}) density profile,

\begin{equation}
\rho(r)=\frac{\rho_{s}}{r/r_{s}(1+r/r_{s})^{2}},
\end{equation}

\noindent where $\rho_{s}$ and $r_{s}$ are the scale density and radius where the logarithmic density slope is equal to $-2$. The ratio between the halo's virial radius $R_{200}$ and the scale radius, $r_{s}$, defines the concentration parameter $c_{200}=R_{200}/r_{s}$ of the profile.

\citet{Green19} provide the best-fit parameters for the transfer function, $H(r,t,f_{b},c_{200})$, defined as the ratio of the evolved subhalo density profile relative to the initial profile, $H(r,t,f_{b},c_{200})=\rho(r,t)/\rho(r,t=0)$, with $f_{b}$ the bound fraction (mass that remains bound to the subhalo while it is tidally stripped) and $c_{200}$ subhalo concentration parameter at $t=0$ (see \citealt{Green19} for more details).

In our model, however, the subhalo DM density profile depends exclusively on the density profile at the previous time-step, not on the initial profile. We therefore assume that the density profile at time-step $t_{n}$, $\rho(r,t_{n})$, is calculated from the density profile at a previous time-step $t_{n-1}$, $\rho(r,t_{n-1})$, via the transfer function as

\begin{equation}\label{ModifiedGreenModel}
\rho(r,t_{n})=\rho(r,t_{n-1})\times H(r,t_{n},f_{b},c_{200}(t_{n-1})),
\end{equation}

\noindent where $H$ depends on the bound fraction, $f_{b}=(1-dM)/M(t_{n})$ defined as the fraction of mass that remains bound after it lost $dM$ mass between $t_{n}$ and $t_{n-1}$, and $c_{200}(t_{n-1})$ the concentration parameter of the density profile at $t_{n-1}$. Although we calculate the amount of mass loss during each $t_{n}$, we do not apply eq.~(\ref{ModifiedGreenModel}) at the end of every time step. This is because the time-step size can become very small, making $dM$ a negligible quantity. Instead, we apply eq.~(\ref{ModifiedGreenModel}) and truncate the density profile every 250 Myr. We have found that during this period of time, the cumulative mass loss of subhaloes reaches on average $1-2\%$ of their total mass. In Appendix \ref{Impact_trunctation_time} we show that truncating the density profile every 350 Myr, instead of 250 Myr, slightly decelerates gravothermal collapse, conversely a more frequent truncation of the density accelerates gravothermal collapse. Section~\ref{Conclusions} discusses the impact of the truncation time parameter on our key results.

Note that eq.~(\ref{ModifiedGreenModel}) is a strong variation of the \citet{Green19} model. We compare the outcome by evolving a $10^{8.84}\Msun$ subhalo that has an initial NFW profile with concentration 15.7, it follows the orbit of UM (shown in top panels in Fig.~\ref{OrbitsDwarfGalaxies}) and loses roughly $40\%$ of its initial mass. We evolve the subhalo using the \citet{Green19} model, as well as the modified model shown in eq.~(\ref{ModifiedGreenModel}). 

\begin{figure} 
	\includegraphics[angle=0,width=0.48\textwidth]{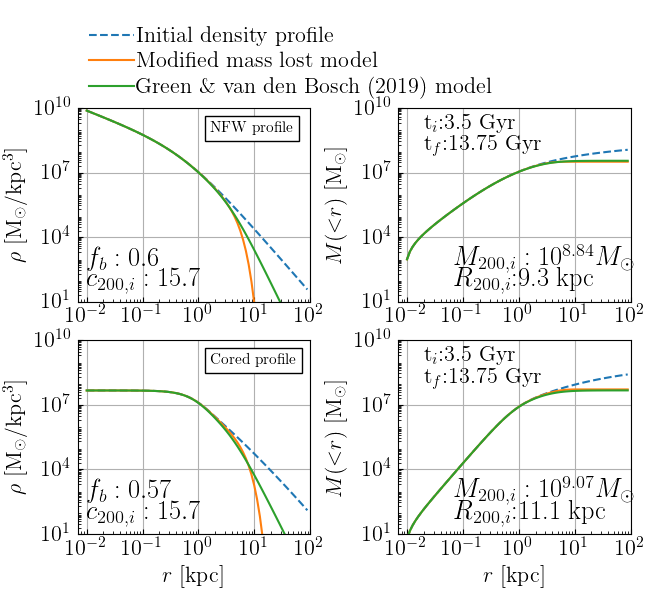}
	\vspace{-0.5cm}
	\caption{Top panels: Density (left) and enclosed mass (right) as a function of radius for a subhalo that has an initial virial mass of $10^{8.84}\Msun$, an initial NFW profile with concentration 15.7, it follows the orbit of UM over 10 Gyr as it loses $40\%$ of its initial mass ($f_{b}=0.6$). The dashed line corresponds to the initial density and mass profile, whereas the solid lines correspond to the final profiles after 10 Gyr. Bottom panels: same as top panels but for a subhalo with an initial virial mass of $10^{9.07}\Msun$ and an initial cored profile. From the left panels it can be seen that the density profiles are largely truncated when we apply eq.~(\ref{ModifiedGreenModel}) in comparison to the Green \& van den Bosch (2019) model, however the large difference only occurs in the outskirts of the halo, beyond the virial radius. The right panels show that the enclosed mass of the two models agree when initializing with NFW-shape or cored-shape profile.}
	\label{TrunctedProfile}
\end{figure}

The top panels of Fig.~\ref{TrunctedProfile} show the density (top-left) and enclosed mass (top-right) as a function of radius of a NFW subhalo. The dashed lines in the panels correspond to the initial density and mass profile, whereas the solid lines correspond to the final profiles after 10 Gyr of evolution. From the top-left panel it can be seen that the density profile is largely truncated when we apply eq.~(\ref{ModifiedGreenModel}) in comparison to the \citet{Green19} model, however the large difference only occurs in the outskirts of the halo, beyond the virial radius. The top-right panel shows that the final masses agree, indicating that the truncation imposed by eq.~(\ref{ModifiedGreenModel}) closely follows the rate of mass loss of the \citet{Green19} model.

An important caveat to consider when we apply eq.~(\ref{ModifiedGreenModel}) is that the transfer function $H$ was fitted according to the structural evolution of an NFW CDM subhalo in the outer regions. Differently, the density profile of SIDM subhaloes largely deviates from the NFW shape in the inner regions, but not in the outer regions, since there the rate of DM-DM particle interactions is low. Given that the transfer function mostly affects the outer regions of the density profile (as shown in the top panels of Fig.~\ref{TrunctedProfile}), we test the evolution of a SIDM subhalo with an initial cored-shape density profile. The bottom panels of Fig.~\ref{TrunctedProfile} show the same as the top but for the SIDM subhalo. After 10 Gyrs of evolution, the subhalo loses roughly $43\%$ of its initial mass, in close agreement with the evolution of the CDM subhalo. Like in the top panels, the density profile is largely truncated when we apply eq.~(\ref{ModifiedGreenModel}) in comparison to the \citet{Green19} model, but mostly beyond the virial radius. 

Given the good agreement between the rate of mass loss and truncated profiles of SIDM and CDM subhaloes in the outer regions, we assume that applying eq.~(\ref{ModifiedGreenModel}) to the truncation of SIDM subhaloes is a good approximation. Note, however, that the rate of mass loss given by eq.~(\ref{TidalStripping}) should be taken as a lower limit. SIDM subhaloes lose more mass due to ram-pressure stripping and the presence of baryons, effects that we do not model in this work. Ram-pressure stripping is caused by DM self-interactions with the host halo particles, that drive material out of subhaloes, to the extend of being able to completely evaporate subhaloes (e.g. see \citealt{Vogelsberger19}). Baryons will not affect the gravothermal SIDM modelling presented in Subsection \ref{Sec_Gravothermal_collapse}, but will enhance the effect of tidal stripping due to the presence of a Galactic disk.

\subsection{Integration of the equations \& initial conditions}\label{Sec_integration_of_eqs}

The gravothermal model comprises eqs.~(\ref{mass_conservation_eq}-\ref{flux_eq_2}) that govern the evolution of the subhalo's density profile given the cross section per unit mass, $\sigma/m_{\chi}$, and the initial subhalo density profile, $\rho_{\rm{init}}$. We set $\rho_{\rm{init}}$ to follow an NFW profile, and divide the spherical subhalo into 150 logarithmically spaced concentric shells, ranging between $r_{\rm{min}}=10^{-2}$ kpc and $r_{\rm{min}}=10^{2}$ kpc.

We solve the gravothermal model by re-writing the equations into non-dimensional form, to do so we introduce a characteristic mass, density and radius, and numerically integrate the equations over a time-step $\Delta t$. For each iteration, $\Delta t$ is restricted to be

\begin{equation}\label{timestep}
\Delta \tilde{t} = {\mathrm{min}}\left[\frac{2}{3}\frac{\tilde{\rho}}{\tilde{v}}(\Delta \tilde{r})^{2}(ab^{-1}\tilde{\sigma_{m}}^{2}+(C\tilde{\rho}\tilde{v}^{2})^{-1})\right],
\end{equation}

\noindent where ``$\sim$" denotes the variables non-dimensional form, and $\Delta r$ the radial spacing of the density profile. We further describe the non-dimensional terms and solution of the gravothermal equations in Appendix A.

We combine the gravothermal model with the orbit integration model, so that at each time-step we calculate the subhalo's distance to the galaxy centre, $d_{GC}$, as well as the instantaneous angular velocity, $\omega$. We use these quantities to determine the amount of mass lost between two consecutive time-steps using eqs.~(\ref{TidalStripping}) and (\ref{TidalRadius}). At each time step we also evolve the MW's virial mass following eq.~(\ref{MAH_model_Correa}), as well as the MW's halo density profile, which we assume to be an NFW profile that follows the concentration-mass relation of the form

\begin{eqnarray}\nonumber
\log_{10}c_{200}(M_{200},z) &=& \alpha(z)+\beta(z)\log_{10}(M_{200}/\Msun)\times\\
&& [1+\gamma(z)(\log_{10}M_{200}/\Msun)^{2}],
\end{eqnarray}

\noindent from \citet{Correa15b}. Every 250 Myrs, we apply eq.~(\ref{ModifiedGreenModel}) in order to truncate the density profile according to the amount of mass lost during that period.

Subhaloes hosting the MW dSphs are initialised with the orbital parameters taken from \citet{Fritz18}, namely the distance to the MW centre, $d_{GC}$, radial velocities, $v_{R}$ and tangential velocities, $v_{T}$, which are used to calculate the orbital evolution of each subhalo. \citet{Fritz18}, as well as \citet{Helmi18}, reported uncertainties in the dwarf distances and radial velocities of the order of $7-8\%$. We have found that such errors do not change the results presented in the following section. The errors of the tangential velocities, however, are larger, of the order of $20\%$ for all dwarfs except CVnI, LeoI and LeoII, which range between 60 and $110\%$. These errors can change the orbits of the dwarfs to a great extent, thus altering the rate of mass loss and affecting the period of gravothermal collapse. In Section~\ref{Uncertainty_orbital_parameters} we discuss how the errors in $v_{T}$ impact on our results.

Besides the orbital parameters, each subhalo is initialised with two free parameters: the cross section per unit mass, $\sigma/m_{\chi}$, and the initial subhalo virial mass, $M_{200,\rm{init}}$. The latter is used to estimate the DM halo concentration parameter at the initial cosmic time $t=3.5$ Gyr (redshift $z=1.87$), using the concentration-mass relation of \citet{Correa15b}, which in turn is used to initialise $\rho_{\rm{init}}$. 

Table \ref{Initial_conditions_table} lists the orbital parameters taken from \citet{Fritz18}, the initial virial mass and set of parameters that describe the initial NFW profile for each subhalo. Note that the initial concentrations for the dSphs are quite low ($c_{200}\sim 6-7$), in opposite to typical $z=0$ values for $10^{9}\Msun$ systems of $c_{200}\sim 15-20$. This is because of two reasons. First $c_{200,\rm{init}}$ is set by assuming that 10 Gyrs ago, before the dSph galaxies became MW's satellites, those galaxies were hosted by field haloes that followed the median concentration-mass relation for $z = 1.87$. Secondly, the concentration-mass relation from \citet{Correa15b} is not a best-fit extrapolation from cosmological simulations, it is a semi-analytic model that combines an analytic model for the halo mass accretion history, based on extended Press Schechter (EPS) theory (\citealt{Press74}), with an empirical relation between concentration and formation time (\citealt{Correa15c}). Because the semi-analytic model is based on EPS theory, it can be applied to wide ranges in mass, redshift and cosmology. Throughout this work we assume Planck13 cosmology (\citealt{Planck}) with $\Omega_{\rm{m}}$, $\Omega_{\Lambda}$, $h$, $\sigma_{8}$, $n_{s}$ equal to 0.307, 0.693, 0.6777, 0.8288, 0.9611, respectively. In Section~\ref{Sec_impact_ICs_2} we discussed how changing the initial values of the concentration parameter impacts on our results.

We run the SIDM halo model for the nine systems hosting the most massive MW dSphs. The evolution begins when the Universe is 3.5 Gyr old, at a point when none of the systems have yet crossed the MW's virial radius, and it finishes at present time, covering 10.2 Gyrs of evolution. In Section~\ref{Sec_impact_ICs_1} we show that using an NFW profile for the initial density profile of subhaloes and MW, rather than a cored profile, does not modify our results.

\begin{table*}
\begin{center}
\begin{tabular}{lrrrcccc}
\hline
\multicolumn{1}{c}{} & \multicolumn{3}{c}{\uline{Orbital parameters}} & \multicolumn{4}{c}{Initial Conditions} \\
\cline{5-8}
Name & $d_{\rm{GC}}$ & $v_{\rm{R}}$ & $v_{\rm{T}}$ & $M_{200,\rm{init}}$ & $c_{200,\rm{init}}$ & $\rho_{\rm{s,init}}$ & $r_{\rm{s,init}}$\\
& [kpc] & [km/s] & [km/s] & [$10^{9}\Msun$] & & [$10^{7}\Msun/$kpc$^{3}$] & [kpc]\\
\hline\hline
UM & 78 & $-71$ & 136 & 0.60 & 6.87 & 1.84 & 1.30\\
Draco & 79 & $-89$ & 134 & 3.46 & 6.36 & 1.54 & 2.52\\
Carina & 105 & 2 & 163 & 2.13 & 6.53 & 1.62 & 2.09\\
Sextans & 89 & 79 & 229 & 0.67 & 6.99 & 1.83 & 1.34\\
CvnI & 211 & 82 & 94 & 1.09 & 6.68 & 1.73 & 1.63s\\
Sculptor & 85 & 75 & 184 & 4.74 & 6.28 & 1.49 & 2.82\\
Fornax & 141 & $-41$ & 132 & 3.54 & 6.38 & 1.53 & 2.54\\
LeoII & 227 & 20 & 74 & 0.14 & 7.30 & 2.13 & 0.76\\
LeoI & 273 & 167 & 72 & 3.23 & 6.40 & 1.55 & 2.44\\
\hline
\end{tabular}
\end{center}
\caption{Form left to right: list of orbital parameters and initial conditions. The first column indicates the name of the dSph galaxy that corresponds to the observational estimates for the galactocentric distance, $d_{\rm{GC}}$, radial and tangential velocities, $v_{\rm{R}}$ and $v_{\rm{T}}$, taken from Fritz et al. (2018). The fifth and sixth columns from the left correspond to the initial virial mass and concentration, $M_{200,\rm{init}}$ and $c_{200,\rm{init}}$, each subhalo is initialised at cosmic time 3.5 Gyr ($z=1.87$) before infalling onto the MW system. The seventh and eighth columns indicate the respective scale density and radius, $\rho_{\rm{s}}$ and $r_{\rm{s}}$, of the initial NFW density profile, $\rho_{\rm{init}}$.}
\label{Initial_conditions_table}
\end{table*}

\section{Results}\label{Results}

In this Section we present the results obtained with the SIDM halo analytic model. We begin by illustrating how the joint framework of gravothermal evolution and gravitational tidal stripping shapes the evolution of the density profile of each system. We next describe the range of values of the free parameters, $\sigma/m_{\chi}$ and $M_{200,\rm{init}}$, that reproduce the central DM densities reported in \citet{Kaplinghat19}. Finally we show that there is a promising range of velocity-dependent cross section models that explain the anti-correlation of central density and pericenter distance for the nine most massive MW dSphs.

\subsection{SIDM halo evolution}\label{Sec_SIDM_halo_evolution}

\begin{figure*} 
	\includegraphics[angle=0,width=0.7\textwidth]{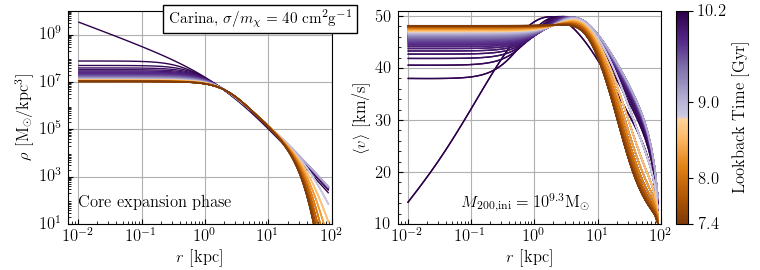}
	\includegraphics[angle=0,width=0.7\textwidth]{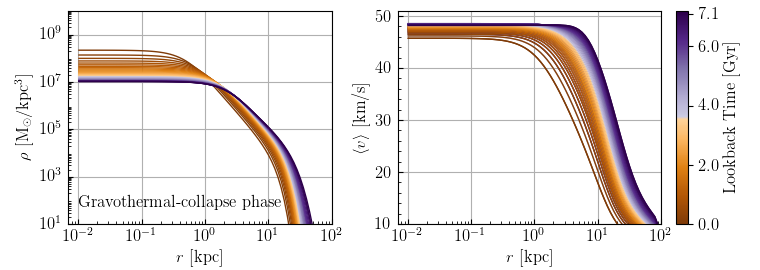}
	\caption{Top panels: Density (left) and velocity profile (right) as a function of radius for the subhalo hosting the galaxy Carina. In this example the model was initialised with a cross section of $\sigma/m_{\chi}=40$ cm$^{2}$g$^{-1}$ and a virial mass of $M_{200}=10^{9.3}\Msun$. The subhalo was evolved for 10.2 Gyrs from an initial NFW profile with scale density and radius of $4.2\times10^{6}\Msun$kpc$^{-3}$ and 2.09 kpc, respectively. Each line in the panels is coloured according to the lookback time, as shown in the colour bar at the top. Bottom panels: same as top panels but for the last 7 Gyrs of evolution, when the system undergoes the gravothermal collapse phase.}
	\label{Result_evol_profile}
\end{figure*}

In this Section we show the evolution of the subhalo that hosts the galaxy Carina. For this system the model was initialised with a cross section of $\sigma/m_{\chi}=40$ cm$^{2}$g$^{-1}$. We assume that 10.2 Gyrs ago (redshift $z=1.87$), Carina had a virial mass of $M_{200}=2\times 10^{9}\Msun$, and its density followed the NFW profile with a scale density and radius of $4.2\times10^{6}\Msun$kpc$^{-3}$ and 2.09 kpc, respectively.

Fig.~\ref{Result_evol_profile} shows the density (left panels) and velocity (right panels) profiles at different times. The velocity corresponds to the particles' average collision velocity, $\langle v\rangle=(4/\sqrt{\pi})v$ (for a Maxwellian distribution), with $v$ the 1-D velocity dispersion. Each line in the figure is coloured according to the lookback time as shown by the colour bars on the right. In the top panels, the dark blue lines correspond to the density and velocity of the system when it begins to evolve, 10.2 Gyrs ago, whereas the orange lines correspond to the evolution of the system between 7.4 and 8.8 Gyrs ago. It can be seen that after a few time-steps the cusp in the central region disappears, the central density decreases and  a core of roughly constant density begins to form.

The top right panel shows that while the density profile rapidly forms a central core, the particles' velocity increases. The density in the inner regions decreases due to DM-DM collisions, that expel some DM particles from the central regions into further out orbits, at the same time the velocity increases because collisions increase the mean velocity of particles.

DM-DM particle collisions add energy to the core, causing particles to move into larger orbits at lower velocities. Through collisions the subhalo's core becomes a system with negative heat capacity, where adding energy cools down the system, while the extended subhalo becomes a large thermal reservoir that absorbs the core's energy. The subhalo stabilises as it forms a high temperature core with negative heat capacity. This is the period between the end of the core expansion phase and the beginning of the gravothermal collapse-phase, that began roughly 9 Gyrs ago for this system and lasts for roughly 3 Gyrs. The bottom panels of Fig.~\ref{Result_evol_profile} show the evolution of the subhalo in the gravothermal-collapse phase. During this phase, the high temperature, negative-heat capacity core in contact with the cold extended halo gives up heat, getting hotter rather than colder. The hot core then contracts, the central density increases, leading to the gravothermal collapse phase. In the case of Carina, the subhalo reaches a stable central density of $10^{7}\Msun$kpc$^{-3}$. During the last 7 Gyrs, as it goes into the gravothermal collapse phase, its density increases an order of magnitude, reaching a value of $2\times 10^{8}\Msun$kpc$^{-3}$. 

An important aspect in the evolution of Carina is the result of the joint gravothermal evolution and gravitational tidal stripping modelling. Differently from previous studies, the contraction phase of the core is not followed by an increase in the particle's velocity as it would be expected. Instead the particles' velocity decreases during the last 4 Gyr of evolution, this is because during this period hydrostatic equilibrium significantly lowers the pressure of the subhalo when it loses mass, causing the velocity to decrease. Since heat flows towards the colder extended halo, heat is diffused at a faster rate when mass is tidally stripped, leading to a faster formation of the isothermal core and thus an accelerated evolution for core collapse.



All systems undergo a similar evolution to the one described in this section, with the only difference that a few systems reach a higher or lower central density during the gravothermal collapse phase, and others lose more or less mass as they orbit around the MW. The following section describes the dependence of central density evolution on the scattering cross section.

\begin{figure*} 
	\includegraphics[angle=0,width=0.75\textwidth]{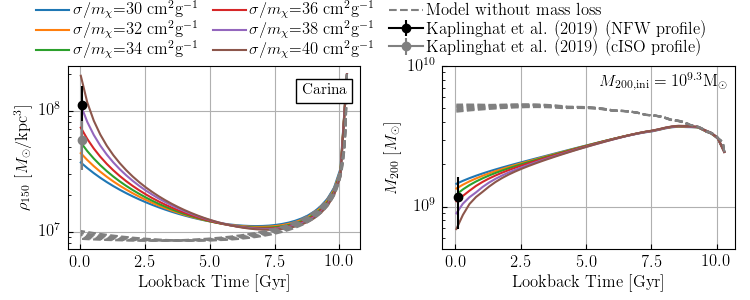}
	\caption{Left panel: Carina's DM density at 150 pc, $\rho_{150}$, as a function of lookback time. The coloured lines correspond to the subhalo model initialised with a different cross section value, ranging from $\sigma/m_{\chi}=32$ cm$^{2}$g$^{-1}$ to $\sigma/m_{\chi}=40$ cm$^{2}$g$^{-1}$, but the same initial virial mass, $M_{200,\rm{init}}=2\times 10^{9}\Msun$. The dashed lines show the evolution of $\rho_{150}$ (and $M_{200}$) in the scenario that the subhalo does not lose mass from tidal interactions. The black symbols show the values of $\rho_{150}$ (and $M_{200}$) taken from Kaplinghat et al. (2019), who assumed an isothermal cored profile as well as NFW. Right panel: same as left panel, but showing the evolution of Carina's virial mass, $M_{200}$, as a function of lookback time.}
	\label{Result_central_density_Carina}
\end{figure*}

\begin{figure*} 
	\includegraphics[angle=0,width=\textwidth]{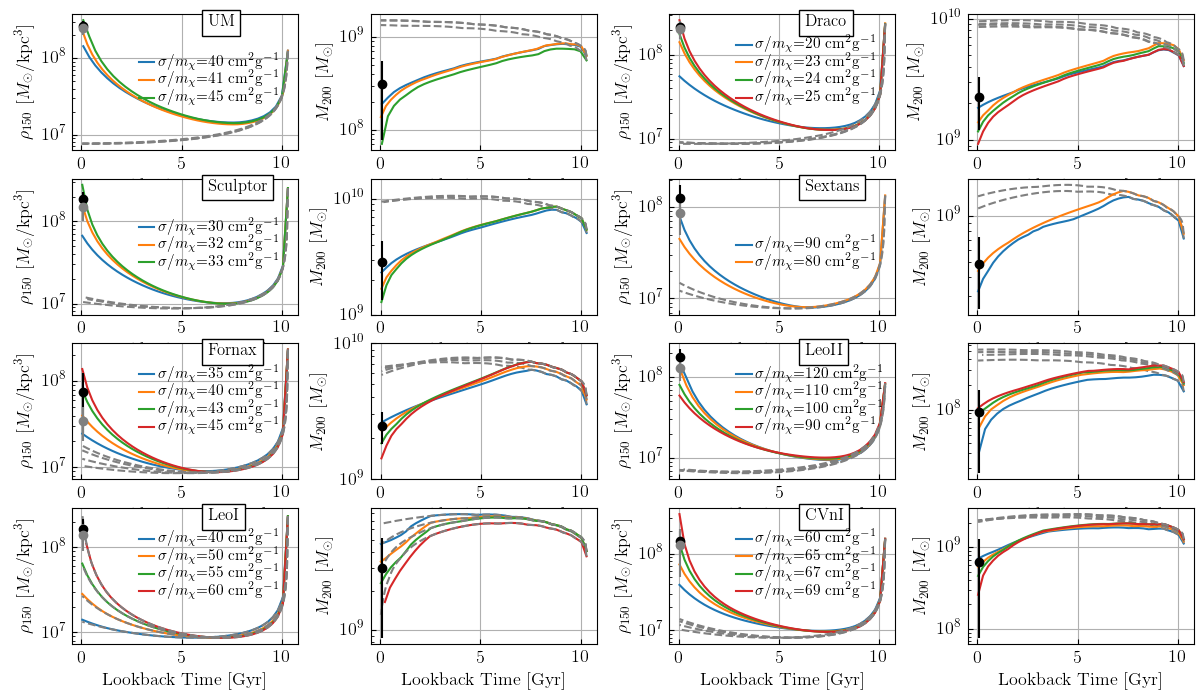}
	\caption{Same as Fig.~\ref{Result_central_density_Carina} for the remaining subhaloes hosting the MW dSphs as indicated in each panel.}
	\label{Result_central_density_all}
\end{figure*}

\subsection{Central density evolution}\label{Sec_central_density_evolution}

The evolution of the central DM density of the subhalo, along with its mass loss rate, largely depends on the scattering cross section. At fixed initial mass, a large cross section leads to a larger rate of DM-DM collisions that produce a shallower and lower density core. Similarly, the larger rate of DM-DM collisions leads to less concentrated subhaloes, making them more prone to tidal disruption and mass loss.

This dependency on the cross section can be seen in Fig.~\ref{Result_central_density_Carina}, that shows the evolution of Carina's DM density at 150 pc, $\rho_{150}$ (left panel), and virial mass, $M_{200}$ (right panel). The coloured lines in the figure correspond to the subhalo model initialised with different values for the cross section, ranging from $\sigma/m_{\chi}=32$ cm$^{2}$g$^{-1}$ to $\sigma/m_{\chi}=40$ cm$^{2}$g$^{-1}$, but the same initial virial mass, $M_{200,\rm{init}}=2\times 10^{9}\Msun$. The dashed lines show the evolution of $\rho_{150}$ and $M_{200}$ without imposing loss of mass from tidal interactions. The black symbols show the values of $\rho_{150}$ and $M_{200}$ reported by \citet{Kaplinghat19}, who assumed both an isothermal cored (grey symbol), as well as NFW (black symbol), profile. We derive $M_{200}$ from the $V_{\rm{max}}$ and $R_{\rm{max}}$ estimations of \citet{Kaplinghat19} assuming an NFW profile for the subhalo density.

The left panel of Fig.~\ref{Result_central_density_Carina} shows that the central density quickly drops when the core of the subhalo forms, and it rises again as the core begins to collapse. For both cases, with or without tidal stripping, the central density reaches a minimum stable value, roughly independent of the cross section. For the model that includes mass loss from tidal stripping, the collapse time becomes shorter than the age of the Universe (as also shown by e.g. \citealt{Nishikawa19}), and the central density reaches higher values for a higher cross section. 

The right panel shows that for the case of no tidal stripping, the subhalo's virial mass slightly increases during its evolution, this is because $M_{200}$ is calculated by integrating the subhalo's density profile until the mean density is $200\times\rho_{\rm{crit}}$, with $\rho_{\rm{crit}}(z)$ decreasing with decreasing lookback time (decreasing redshift). For the model with tidal stripping, the panel shows that during the last 2.5 Gyr, the rate of mass loss is higher for higher cross sections.

Fig.~\ref{Result_central_density_Carina} shows that cross sections ranging between $\sigma/m_{\chi}=32$ cm$^{2}$g$^{-1}$ and $\sigma/m_{\chi}=40$ cm$^{2}$g$^{-1}$ are able to explain the observed central DM density of Carina (within the uncertainty), as well as the final virial mass of the system. However, this range of parameters seems to only apply to Carina. Fig.~\ref{Result_central_density_all} shows the evolution of DM density at 150 pc and virial mass for the remaining subhaloes hosting the MW dSphs: UM, Draco, Sculptor, Sextans, Fornax, LeoII, LeoI and CVnI. From the figure it can be seen that there is a large variety of cross sections that reproduce the observed central densities. Draco, for instance, prefers lower values of $\sigma/m_{\chi}$, ranging between $\sigma/m_{\chi}=23$ cm$^{2}$g$^{-1}$ and $\sigma/m_{\chi}=25$ cm$^{2}$g$^{-1}$, whereas LeoII requires $\sigma/m_{\chi}$ ranging between $\sigma/m_{\chi}=120$ cm$^{2}$g$^{-1}$ and $\sigma/m_{\chi}=90$ cm$^{2}$g$^{-1}$.

An interesting case shown in Fig.~\ref{Result_central_density_all} is that of LeoI, since this subhalo crossed the MW's virial radius roughly 2 Gyrs ago, it has not lost a large amount of mass from tidal interactions with the MW. Therefore the models with and without mass loss agree.

Fig.~\ref{Result_central_density_all} shows that large cross sections are needed in order to reproduce the observed central DM densities of the local dSphs. Interestingly, such large cross sections are not ruled out by observational constraints related to the TBTF. \citet{Vogelsberger12} produced zoom simulations of MW-size hosts using SIDM with velocity-dependent $\sigma/m_{\chi}$ values tuned to have small values ($\sim 0.1$ cm$^{2}$g$^{-1}$) on cluster scales and large values ($\sim 10$ cm$^{2}$g$^{-1}$) on the scale of dwarf galaxies. They showed that the velocity-dependent SIDM model resolved the TBTF problem, and it provided a particular good match to the spread in dwarf satellite central densities seen around the MW. \citet{Elbert15} showed that even larger cross sections, e.g. $\sigma/m_{\chi}=50$ cm$^{2}$g$^{-1}$, also alleviate the TBTF problem and produce constant density cores of size 300-1000 pc, comparable to the half-light radii of $\sim 10^{5-7}\Msun$ stellar mass dwarfs. 


Table~\ref{Final_profile_table} summarises the final density profiles for the SIDM subhaloes presented in this section. From left to right, it shows the hosted dSph galaxy name, virial mass, concentration parameter, core size and the preferred range of cross section values. Note that the virial mass is calculated by integrating the density profile up to the virial radius, which in turn is estimated as the radius within which the mean density is 200 times $\rho_{\rm{crit}}(z=0)$. The core radius, $r_{\rm{c}}$, is calculated by fitting an isothermal profile ($\rho(r)=\rho_{0}/(r_{\rm{c}}^{2}+r^{2})$).

Figs.~\ref{Result_central_density_Carina} and \ref{Result_central_density_all} show that all MW dSphs need to be in gravothermal core collapse in order to explain the observational data. This result, however, strongly depends on the initial virial mass of the systems, $M_{200,\rm{init}}$, which is not chosen at random, it is tuned so that the systems, in their final state, have a virial mass that reproduces the observational estimations. If we disregard this and increase the initial mass of the systems, lower values of $\sigma/m_{\chi}$ will be needed to reproduce the observed central DM densities. But again, they would not be a good theoretical representation of the dSphs because, even considering that the model's rate of mass loss is a lower limit, the systems would too massive. 

In Appendix~\ref{Comparison_Minit_appendix} we show that changing $M_{200,\rm{init}}$ in $20\%$ can change the final central DM densities of subhaloes in up to $50\%$. Although the final DM central density is quite sensitive to the choice of initial mass, it also largely depends on $\sigma/m_{\chi}$. Fig.~\ref{Result_central_density_Carina} shows that for a constant $M_{200,\rm{init}}$, changes in $\sigma/m_{\chi}$ of up to $25\%$ lead to changes of $80\%$ in the final central DM densities.

\begin{table}
\begin{center}
\begin{tabular}{lccrc}
\hline
\multicolumn{1}{c}{} &  \multicolumn{3}{c}{Final profile} & \multicolumn{1}{c}{\uline{Preferred cross section}}\\
\cline{2-4}
Name & $M_{200}$ & $c_{200}$ & $r_{\rm{core}}$ & $\sigma/m_{\chi}$ \\
&  [$10^{9}\Msun$] & & [pc] & [cm$^{2}$g$^{-1}$] \\
\hline\hline
UM & 0.13 & 34.2 & 180.8 & $40-50$\\
Draco & 1.17 & 26.8 & 472.9 & $20-30$\\
Carina & 1.09 & 19.1 & 648.4 &  $40-50$\\
Sextans & 0.32 & 20.8 & 395.5 & $70-120$\\
CVnI & 0.46 & 25.7 & 356.8 & $50-80$\\
Sculptor & 1.65 & 25.8 & 553.2 & $30-40$\\
Fornax & 2.29 & 15.3 & 1036.7 & $30-50$\\
LeoII & 0.05 & 30.6 & 148.8 & $90-150$\\
LeoI & 1.17 & 31.1 & 410.8 & $50-70$\\
\hline
\end{tabular}
\end{center}
\caption{Form left to right: name of the dSph galaxy, present-time virial mass, concentration parameter and core size of the subhalo hosting the dSph and range of preferred cross section values that reproduce the observed DM central densities. }
\label{Final_profile_table}
\end{table}

\begin{figure} 
	\includegraphics[angle=0,width=0.48\textwidth]{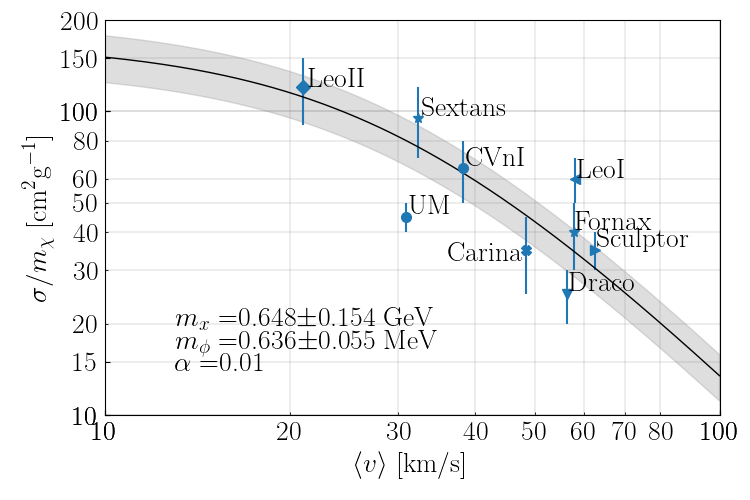}
	\caption{Cross section per unit mass, $\sigma/m_{\chi}$, as a function of the average collision velocity, $\langle v\rangle$, of DM particles within each subhalo's core. Symbols show the range of $\sigma/m_{\chi}$ needed for the SIDM model to reproduce the central DM densities reported by Kaplinghat et al. (2019). The solid line corresponds to the best-fit relation given by eq.~(\ref{cross_section_total}) to the MW dSph data.}
	\label{Velocity_cross_section_plane}
\end{figure}

\subsection{Velocity-dependent cross section}\label{Sec_Velocity_cross_section}

In this section we analyse the range of cross sections that match the observed central DM densities shown in Figs.~\ref{Result_central_density_Carina} and \ref{Result_central_density_all}. To understand its dependence with the particles velocity we calculate the average collision velocities, $\langle v\rangle$, of DM particles within each system's core. We define $\langle v\rangle=(4/\sqrt{\pi})v$ (for a Maxwellian distribution), with $v$ the average 1-D velocity dispersion of each system's core. We find that for the case of Carina for example (Figs.~\ref{Result_evol_profile} and \ref{Result_central_density_Carina}), observations favour a cross section range between 32 and 40 cm$^{2}$g$^{-1}$, as its core reaches a stable collision velocity of $\approx 48$ km/s. 

We find that there is a strong correlation between $\sigma/m_{\chi}$ and $\langle v\rangle$. This is shown in Fig.~\ref{Velocity_cross_section_plane}, that highlights the range of values for each dSph. From the figure it can be seen that for systems such as LeoII, characterised by $\langle v\rangle\approx 21$ km/s, observations favour a large cross section, whereas for Draco, which has a $\langle v\rangle\approx$58 km/s, observations favour lower cross sections. We determine the range of cross sections for each dSph by analysing the models with fixed $\sigma/m_{\chi}$ that are able to reproduce the DM central densities (Fig.~\ref{Result_central_density_all}), and also by considering that DM densities from the gravothermal model may differ from N-body simulations by up to a factor of 2 (\citealt{Essig19}). We used this uncertainty to further extend the range of $\sigma/m_{\chi}$.

Fig.~\ref{Velocity_cross_section_plane} indicates that the range of cross sections needed to reproduce the observed DM densities are not random, instead they point towards an intrinsic velocity-dependent relation. We investigate this relation in the context of particle physics models for SIDM, where a DM particle $\chi$ of mass $m_{\chi}$ interacts under the exchange of a light mediator $\phi$, with the self-interactions being described by a Yukawa potential

\begin{equation}\label{Yukawa_potential}
V(r)=\pm \frac{\alpha_{\chi}}{r}e^{-m_{\phi}/r},
\end{equation}

\noindent with $r$ the separation between DM particles, $\alpha_{\chi}$ the analog of the fine-structure constant in the dark sector, and $m_{\phi}$ the mediator mass. 

There is no analytical form for the differential scattering cross section due to a Yukawa potential, but by using the Born-approximation (valid when the scattering potential can be treated as a small perturbation), the analytical form that approximates the true differential cross section results (see e.g. \citealt{Ibe10,Tulin13,Tulin18}) 

\begin{equation}\label{cross_section_angle}
\frac{{\rm{d}}\sigma}{{\rm{d}}\Omega}=\frac{\sigma_{0}}{4\pi}\left[1+\frac{v^{2}}{w^{2}}\sin^{2}\left(\frac{\theta}{2}\right)\right]^{-2},
\end{equation}

\noindent where $v$ is the relative velocity between interacting DM particles, $w\equiv m_{\phi}/m_{\chi}$ is a characteristic velocity, $\sigma_{0}=4\pi \alpha_{\chi}^{2}m_{\chi}^{2}/m_{\phi}^{4}$ is the amplitude of the cross section, and $\theta$ is the scattering angle in the frame of the centre of mass. From eq.~(\ref{cross_section_angle}) we calculate the total cross section by integrating over the solid angle and obtain

\begin{eqnarray}\nonumber
\sigma/m_{\chi} &=& 0.0275\left(\frac{\alpha_{\chi}}{0.01}\right)^{2}\left(\frac{m_{\chi}}{10\,{\rm{GeV}}}\right)\left(\frac{10\,{\rm{MeV}}}{m_{\phi}}\right)^{4}\\\label{cross_section_total}
&& \times\,\left(1+v^{2}/w^{2}\right)^{-1}{\rm{cm}}^{2}{\rm{g}}^{-1},\\\nonumber
\rm{with} &&\\\nonumber
w & = & 30\left(\frac{m_{\phi}}{10\,{\rm{MeV}}}\right)\left(\frac{10\,{\rm{GeV}}}{m_{\chi}}\right)\,\rm{km}\,\rm{s}^{-1}.
\end{eqnarray}

We fit eq.~(\ref{cross_section_total}) to the dSphs values in order to determine the values of DM mass and mediator mass that reproduce the relation. We assume $\alpha_{\chi}=0.01$ and find that the relation is best fitted by a DM particle mass of $m_{\chi}=0.648\pm 0.154$ GeV and a mediator mass of $m_{\phi}=0.636\pm 0.055$ MeV. This best-fit relation is shown in Fig.~\ref{Velocity_cross_section_plane} in solid line, the grey region highlights the uncertainty by propagating the errors of $m_{\chi}$ and $m_{\phi}$.

\begin{figure} 
	\includegraphics[angle=0,width=0.48\textwidth]{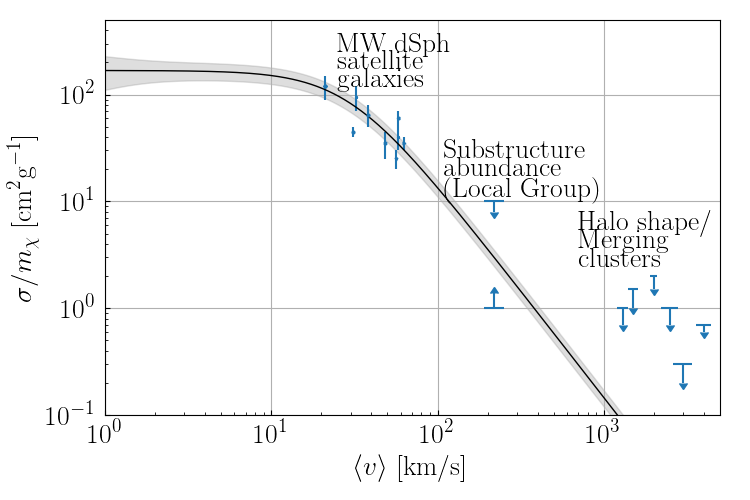}
	\caption{Same as Fig.~\ref{Velocity_cross_section_plane}, but extended to cover the range of MW- (${\sim}200$ km/s) and cluster-size (${\sim}1000-5000$ km/s) haloes' velocities. The figure shows upper and lower limits for $\sigma/m_{\chi}$ taken for substructure abundance studies (e.g. Volgelsberger et al. 2012 and Zavala et al. 2013), as well as based on halo shape/ellipticity studies and cluster lensing surveys (see text).}
	\label{Velocity_cross_section_plane_ALL}
\end{figure}

Fig.~\ref{Velocity_cross_section_plane_ALL} extends the cross section-velocity plane to include MW- ($v\sim 150-300$ km$/$s) and cluster-size ($v\sim 1000-5000$ km$/$s) haloes. In the figure, the values for the MW dSph galaxies are shown in blue symbols and the best-fit relation in solid line. The figure shows that the extrapolation of the best-fit cross section-velocity relation to large scales lies in very good agreement with current observational constraints. For characteristic DM velocities of MW-size haloes, the upper limit of $\sigma/m_{\chi}<10$ cm$^{2}$g$^{-1}$ is set by subhalo evaporation, where collisions between DM particles within subhaloes and in the host are frequent enough to unbind material from the halo. In this case, energy transfer is determined by the relative velocity of the colliding particles, which is of the order of the orbital velocity, therefore the mass loss in subhaloes is enhanced and the subhalo abundance is depleted relative to the CDM case, particularly in the central regions (e.g. \citealt{Vogelsberger12,Rocha13,Zavala13}). Note that the large values of $\sigma/m_{\chi}$ on dwarf-scales does not translate into evaporation of substructure, this is because the relative velocity between the subhalo and its host halo is set by the velocity dispersion of the latter, for which the cross section is suppressed.

The lower limit of $\sigma/m_{\chi}>1$ cm$^{2}$g$^{-1}$ has been imposed to solve the cusp-core and TBTF problem, otherwise dwarfs and low surface brightness galaxies in SIDM models end up too dense and do not produce large enough core radii (e.g. \citealt{Dave01,Kaplinghat16}). For $\sigma/m_{\chi}=1$ cm$^{2}$g$^{-1}$, \citet{Robles19} find very similar substructure abundance around the MW between the SIDM model and CDM. Differently, \citet{Nadler20} performed high-resolution zoom-in simulations of a MW-mass halo and reported that $56\%$ was disrupted and erased due to subhalo-host halo interactions for a $\sigma/m_{\chi}=2$ cm$^{2}$g$^{-1}$. Neither of these works included baryons in their simulations, but \citet{Robles19} embedded subhaloes in a baryonic potential to capture effects of the disk and bulge contributions.

For characteristic DM velocities of cluster-size haloes, the upper limits in the figure correspond to strong lensing measurements of the ellipticity and central density of clusters (\citealt{Peter13}), measurements of the offset between the DM and galaxy centre (e.g. \citealt{Kahlhoefer15,Harvey15,Wittman18,Harvey19}) and measurements of the mass-to-light ratio of the Bullet Cluster (\citealt{Randall08}).

\section{Discussion}\label{Sec_Discussion}

\subsection{Comparison with previous works}

Recent works have also investigated the evolution of MW dSph galaxies aiming to constrain the SIDM cross section. \citet{Read18} and \citet{Valli18} considered a subhalo's inner region (limited by a radius $r_{\chi}$), within which the average scattering rate per particle times the halo age ($t_{\rm{age}}$) is equal to unity. At $r=r_{\chi}$, therefore, the cross section can be constrained from the relation

\begin{equation}\label{sigma_cored_model}
\sigma/m_{\chi}\simeq (\sqrt{\pi}/4)\rho(r_{\chi})^{-1}v(r_{\chi})^{-1}t_{\rm{age}}^{-1},
\end{equation}

\noindent where $\rho(r)$ is the density profile and $v(r)$ the velocity dispersion. 

\citet{Valli18} analysed the stellar kinematic dataset for the MW dSphs, applying the standard Jeans analysis, to infer the stellar and DM density and velocity dispersion. They assumed $t_{\rm{age}}$ to be flatly distributed in the range 8-12 Gyr and found that UM, Draco, LeoI and LeoII probed cross sections $\sim 0.1-1$ cm$^{2}$g$^{-1}$, whereas Sextans and Fornax had cross section values that peaked around $\sim 20-40$ cm$^{2}$g$^{-1}$. Sculptor and Carina probed intermediate cross section values of $\sim 2-7$ cm$^{2}$g$^{-1}$.

\citet{Read18} focused on Draco and first calibrated the parameters $r_{\chi}$ and $t_{\rm{age}}$ using \citet{Vogelsberger12} SIDM cosmological zoom simulations of MW-mass haloes. They found that Draco's high central density gives an upper bound on the SIDM cross section of $\sigma/m_{\chi}< 0.57$ cm$^{2}$g$^{-1}$. 

The model given by eq.~(\ref{sigma_cored_model}) does not include the effects of core collapse, we test it by applying eq.~(\ref{sigma_cored_model}) to our simulated subhaloes assuming $t_{\rm{age}}$ in the range 8-12 Gyr, and never recover a $\sigma/m_{\chi}$ larger than 1 cm$^{2}$g$^{-1}$. In addition, the model is only valid if the SIDM subhalo density profile follows a cored profile. Observational studies, however, have reported that only Fornax exhibits a prominent core (e.g. \citealt{Jardel12,Pascale18}), Draco (e.g. \citealt{Read18}) and the remaining MW dSphs are better described by a cuspy profile (e.g. \citealt{Read19} and references therein). Interestingly, for Fornax, \citet{Valli18} reported a $\sigma/m_{\chi}\approx 40$ cm$^{2}$g$^{-1}$, in agreement with our results

\subsection{Challenges}\label{Sec_Caveats}

An important caveat of the gravothermal collapse model is that the parameter $C$ cannot be derived from first principles, instead it needs to be calibrated using N-body simulations. Previous works have done it (\citealt{Balberg02,Koda11,Essig19,Nishikawa19}), but have not reach to a consensus of its value, other than it ranges between 0.45 and 0.75. In this section we investigate how changing $C$ from 0.75 (assumed so far) to 0.45 or 0.6 (and $b=0.003$ as suggested by \citealt{Nishikawa19}) impact on our results.

\begin{figure*} 
	\includegraphics[angle=0,width=0.75\textwidth]{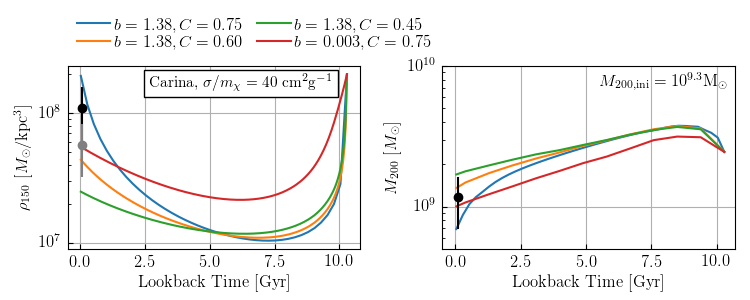}
	\caption{Left panel: Carina's DM density at 150 pc, $\rho_{150}$, as a function of lookback time. The coloured lines correspond to the subhalo model initialised with a cross section of $\sigma/m_{\chi}=40$ cm$^{2}$g$^{-1}$, but different values for the parameters $C$ and $b$ that govern the gravothermal collapse model (see Section~\ref{Sec_Gravothermal_collapse} for further details). The symbols show the values of $\rho_{150}$ (and $M_{200}$) taken from Kaplinghat et al. (2019), who assumed an isothermal cored profile (grey symbol) as well as NFW (black symbol). Right panel: same as left panel, but showing the evolution of Carina's virial mass, $M_{200}$, as a function of lookback time. The figure shows that the parameters $C$ and $b$ impact on our results but not by a large factor.}
	\label{Comparison_C_cte}
\end{figure*}

Fig.~\ref{Comparison_C_cte} shows the DM density at 150 pc, $\rho_{150}$ (left panel), and virial mass, $M_{200}$ (right panel), of a subhalo hosting a Carina-like dSph galaxy. The coloured lines correspond to the subhalo model initialised with a cross section of $\sigma/m_{\chi}=40$ cm$^{2}$g$^{-1}$ and same initial profile, but different values for the parameters $C$ and $b$.

For all models, a cross section of $\sigma/m_{\chi}=40$ cm$^{2}$g$^{-1}$ yields close agreement with the estimations from \citet{Kaplinghat19} (shown as grey and black symbols). For a fixed cross section and $b=1.38$, a larger $C$ accelerates core collapse. In this manner the figure indicates that the range of cross sections (presented in Section~\ref{Results}), derived with a model that assumes $C=0.75$, should be taken as a lower limit. A model with $b=1.38$ and $C=0.45$ requires a factor of ${\approx}1.5$ larger cross sections to reproduce the observed central DM densities and virial masses.

The model with $C=0.75$ and $b=0.003$ largely differs from the rest. It does not lower the central DM density as much during the core expansion phase, and the rate at which the central DM density increases during the core collapse phase seems to agree with the model with $b=1.38$ and $C=0.45$. For a fixed cross section and $C=0.75$, a lower $b$ yields lower central DM densities and a `slower' core collapse in comparison to the $C=0.75$ and $b=1.38$ model. Therefore, according to this model, the range of cross sections presented in Section~\ref{Results} should also be taken as lower limits, since assuming $b=0.003$ would result in larger values of $\sigma/m_{\chi}$ being able to reproduce the observations.

Another important caveat to consider is the assumption of mass conservation in the gravothermal collapse model (eq.~\ref{mass_conservation_eq}), that is inconsistent with the fact that the subhalo loses mass every 250 Myrs. To solve this we impose that after the subhalo loses mass, it returns to hydrostatic equilibrium, readjusting its density and pressure within each shell. After that the gravothermal model is called back, and the new evolution of the truncated density profile begins. Further details of the numerical implementation of the gravothermal model are presented in Appendix~\ref{Gravothermal_model_appendix}. 


\subsection{Impact of initial conditions: NFW profile}\label{Sec_impact_ICs_1}

\begin{figure*} 
	\includegraphics[angle=0,width=0.8\textwidth]{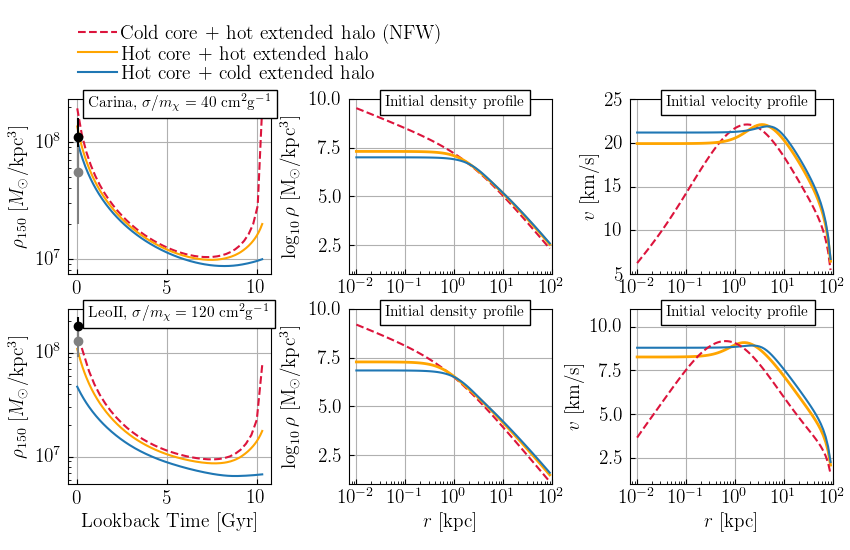}
	\caption{DM density at 150 pc, $\rho_{150}$, as a function of lookback time (left), initial density profile (middle) and initial 1-D velocity dispersion profile (right) for Carina (top panels) and Leo II (bottom panels). Carina was initialized with a cross section of $\sigma/m_{\chi}=40$ cm$^{2}$g$^{-1}$, whereas Leo II has a cross section of $\sigma/m_{\chi}=120$ cm$^{2}$g$^{-1}$. The coloured lines correspond to the subhalo model initialised with a NFW density profile (red dashed lines), cored profile with a hot extended halo (yellow solid lines) and cored profile with a cold extended halo (blue solid lines). The symbols show the values of $\rho_{150}$ taken from Kaplinghat et al. (2019), who assumed an isothermal cored profile (grey symbol) as well as NFW (black symbol). The figure shows that initializing the subhaloes' density profile with either an NFW profile or cored profile (with hot extended halo) does not impact the results presented in Section~\ref{Results}, however initializing the subhaloes' density with a cored profile surrounded by a hot extended halo slightly alters the evolution of subhaloes.}
	\label{Comparison_Initial_Profile}
\end{figure*}

The SIDM halo model presented in this work evolves the subhaloes hosting the most massive MW dSphs for 10.2 Gyrs, starting when the Universe is 3.5 Gyr old (redshift 1.87) at a point when the subhaloes' initial density ($\rho_{\rm{init}}$) is assumed to follow the NFW profile. This, however, may not be a good assumption. \citet{Harvey18} analysed the evolution of 19 low-mass dwarf spheroidal galaxies using a SIDM numerical simulation with $\sigma/m_{\chi}=10$ cm$^{2}$g$^{-1}$, finding that the dwarf galaxies were already forming a core within the first 2 Gyrs of cosmic time.

It is possible that initializing subhaloes with a cored profile will induce an earlier gravothermal collapse, that will yield lower estimates for the cross sections with respect to the ones reported in Section~\ref{Sec_Velocity_cross_section}. In this section we investigate if this occurs by analysing the evolution of Carina and Leo II, that were modelled with $\sigma/m_{\chi}=40$ cm$^{2}$g$^{-1}$ and $\sigma/m_{\chi}=120$ cm$^{2}$g$^{-1}$, respectively, and initialized with three different density profiles.

The middle panels of Fig.~\ref{Comparison_Initial_Profile} show the initial density profile for the models of Carina (top) and Leo II (bottom). In the panels the red dashed lines correspond to the NFW density profile, characterized by a cuspy and cold inner region surrounded by a hot extended halo. The solid lines correspond to two different core profiles, one where the inner region is hot and it is surrounded by a hot extended halo (yellow solid lines), and the other where the extended halo is cold (blue solid lines). This can be seen in the right panels that show the 1-D velocity dispersion as a function of radius. The difference between these profiles is that they correspond to different evolutionary stages of SIDM haloes. The hot core + hot extended halo corresponds to a SIDM halo that has a hot inner region due to DM-DM particle interactions, but has a hotter periphery due to dynamical heating, induced by mergers and large DM accretion (\citealt{Colin02}). On the contrary, the hot core + cold extended halo corresponds to a SIDM halo that has been in isolation.

The left panels of Fig.~\ref{Comparison_Initial_Profile} show the evolution of the DM density at 150 pc, $\rho_{150}$, for Carina (top) and Leo II (bottom). It can be seen from the top-left panel that initializing Carina with either an NFW profile or cored profile does not impact the results presented in Section~\ref{Results}. Differently, the bottom-right panel shows that initializing Leo II with a core profile of hot inner region and cold extended halo, changes the evolution $\rho_{150}$, reaching lower values at present time. We find, however, that the `hot core + hot extended halo' profile better represents the initial stages of subhaloes hosting the MW dSphs. When the Universe is 3.5 Gyr old, is very likely that the dSph subhaloes have undergone recent mergers or had large rates of mass accretion, since at that point none of them have yet crossed the MW's virial radius. 

Another important assumption of the SIDM halo model is the NFW profile for the MW halo. We find, however, that this is a good approximation for our models. At MW halo scale, $\sigma/m_{\chi}\sim 1-5$ cm$^{2}$g$^{-1}$, according to the $\sigma/m_{\chi}$-velocity relation shown in Fig.~\ref{Velocity_cross_section_plane_ALL}. For these cross sections, \citet{Robles19} has shown that the SIDM MW halo embedded in a baryonic potential not only exhibits a remarkably similar density profile to that of a CDM MW-like halo, but it also has no discernible core.

\subsection{Impact of initial conditions: halo concentration}\label{Sec_impact_ICs_2}

It has previously been shown that the core collapse time-scale, $t_{\rm{c}}$, is (\citealt{Kaplinghat19,Essig19,Nishikawa19})

\begin{equation}\label{time_scale_collapse}
t_{\rm{c}}\propto C^{-1}(\sigma/m_{\chi})^{-1}M_{200}^{-1/3}c_{200}^{-7/2},
\end{equation}

\noindent where $C$ is the free parameter described in Subsection~\ref{Sec_Caveats}. Eq.~(\ref{time_scale_collapse}) indicates that for fixed $\sigma/m_{\chi}$, $c_{200,\rm{init}}$ and $M_{200,\rm{init}}$, the larger parameter $C$ accelerates core collapse as shown in Fig.~\ref{Comparison_C_cte}. It also indicates that a larger cross section and/or virial mass accelerates core collapse. Section \ref{Sec_central_density_evolution} comments that in our model, the initial virial mass of the systems is constrained by the tidal evolution model and observations. This leaves us wondering about the impact of the concentration parameter, $c_{200,\rm{init}}$, on the SIDM subhalo evolution.

\citet{Sameie20} explored the tidal evolution of SIDM subhaloes in the MW's tides. They produced N-body simulations of dwarf spheroidal galaxies orbiting around the MW, modelled with a static potential that included both the disk and bulge components. They found that a constant cross section of $\sigma/m_{\chi}=3$ cm$^{2}$g$^{-1}$ can reproduce the observed DM density profiles of the MW dSphs Draco and Fornax. However, this was only possible if subhaloes were initialised with a large concentration parameter, such as 29.5 for Draco. They showed that if the concentration parameter was lowered to 22.9, not even the model of $\sigma/m_{\chi}=10$ cm$^{2}$g$^{-1}$ was able to reproduce the large DM densities of Draco, and a model with higher cross section was needed (see also \citealt{Kahlhoefer19}).

We test the impact of $c_{200,\rm{init}}$ by running the models of Draco and Fornax initialized with a concentration of $c_{200,\rm{init}}=15$ and $\sigma/m_{\chi}=3$ cm$^{2}$g$^{-1}$. We find that both models reproduce the observed central DM densities, in agreement with \citet{Sameie20}. This result is presented in Appendix~\ref{Comparison_c200_appendix}. 

We believe, however, that setting such large initial concentrations is not well justified. In the starting point of our model, subhaloes represent typical $z=1.87$ low-mass subhaloes in the field, whose density profiles follow the median $z=1.87$ concentration-mass relation. \citet{Correa15b} showed that at high redshift ($z>1$), the halo has large rates of accretion, with its mass history mainly characterized by an exponential growth. During this time, the scale radius increases simultaneously with the virial radius, hence the concentration hardly grows. At low redshift ($z<1$), there is a drop in the accretion and merger rates of small haloes, and the halo mass increases due to the evolution of the reference density used in the spherical overdensity definition of the halo ($\rho_{\rm{crit}}$ in this case, also referred as pseudo-evolution phase). This leads subhaloes to have roughly constant scale radius but growing virial radius, which produces the rapid growth of concentrations. In our initial conditions, subhaloes have not reached the pseudo-evolution phase, therefore their concentrations should be set to low values. 

\subsection{Uncertainty in orbital parameters}\label{Uncertainty_orbital_parameters}

\begin{figure*} 
\begin{center}
	\includegraphics[angle=0,width=0.8\textwidth]{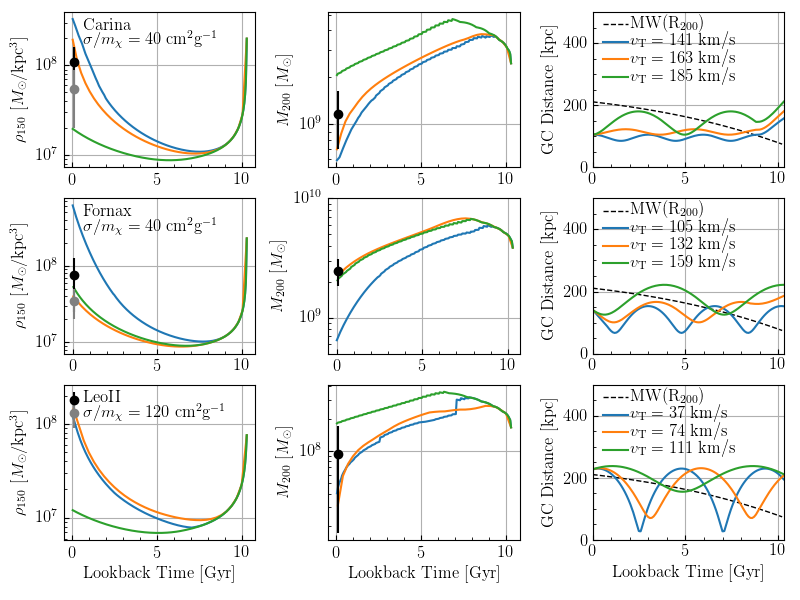}
	\caption{DM density at 150 pc, $\rho_{150}$ (left panels), virial mass, $M_{200}$ (middle panels), and galactocentric distance (right panels) as a function of lookback time for Carina (top panels), Fornax (middle panels) and LeoII (bottom panels). The coloured lines correspond to the subhalo models initialised with the same cross section (of $\sigma/m_{\chi}=40$ cm$^{2}$g$^{-1}$ for Carina and Fornax, and $\sigma/m_{\chi}=120$ cm$^{2}$g$^{-1}$ for LeoII), initial mass, galactocentric distance and radial velocity, but different tangential velocities, $v_{T}$. The symbols show the values of $\rho_{150}$ (and $M_{200}$) taken from Kaplinghat et al. (2019), who assumed an isothermal cored profile (grey symbol) as well as NFW (black symbol). The dashed lines in the right panels indicate the MW virial radius. The figure shows that changes in the tangential velocity of $20\%$ and $50\%$ can largely impact on the final subhaloes DM density at 150 pc.}
	\label{Impact_vT_fig}
	\end{center}
\end{figure*}

This work analyses whether the anti-correlation between the central DM density of MW dSphs, $\rho_{150pc}$, and their pericenter passages, $r_{P}$ (\citealt{Kaplinghat19}), is the result of SIDM effects. The errors in the orbital parameters reported by the Gaia collaboration (\citealt{Fritz18,Helmi18,Brown18}), however, can weaken the anti-correlation. The parameter that produces the largest uncertainties in the pericenter distances is the tangential velocity, $v_{T}$, whose errors are around $20\%$ for Carina, UM, Draco, Fornax, Sculptor and Sextans, but increase to $60-100\%$ for CVnI, LeoI and LeoII. 

Like the $\rho_{150pc}-r_{P}$ anti-correlation, the cross section-velocity relation obtained in Section~\ref{Sec_Velocity_cross_section} can be affected by the uncertainties of $v_{T}$. In this section we investigate this further by analysing how the central DM densities of the dSphs Carina, Fornax and LeoII depend on $20\%$ ($50\%$ for LeoII) changes in the tangential velocity.

Fig.~\ref{Impact_vT_fig} shows $\rho_{150}$ (left panels), virial mass, $M_{200}$ (middle panels) and galactocentric distance (right panels) as a function of lookback time, for Carina (top panels), Fornax (middle panels) and LeoII (bottom panels). The coloured lines correspond to the subhalo models initialised with the same cross section, initial mass, galactocentric distance and radial velocity, but different tangential velocities. The default values of $v_{T}$ for Carina, Fornax and LeoII are 163, 132 and 74 km $\rm{s}^{-1}$, respectively, these are shown in orange solid lines. Blue solid lines indicate the evolution of the models with $20\%$ ($50\%$) larger $v_{T}$ than the default models, whereas green solid lines show the evolution of the models with $20\%$ ($50\%$) lower $v_{T}$ than the default models. 

The top panel shows that lowering $v_{T}$ in $20\%$, decreases the galactocentric distance, increasing the rate of mass loss and accelerating the gravothermal collapse. As a result, Carina reaches a $\sim 50\%$ higher $\rho_{150}$ at present time with respect to the default model. Similarly, increasing $v_{T}$ in $20\%$, increases the galactocentric distance, decreases the rate of mass loss, and Carina reaches a $\sim 60\%$ lower $\rho_{150}$. 

For the evolution of Fornax, lowering $v_{T}$ in $20\%$ yields an earlier infall onto the MW gravitational potential, further increasing the rate of mass loss and accelerating the gravothermal collapse. For this case Fornax reaches over an order of magnitude higher $\rho_{150}$ at present time with respect to the default model. Increasing $v_{T}$, on the other hand, results in a final central DM density in close agreement with the default model.

The changes in $v_{T}$ for LeoII are of the order of $50\%$. The bottom panels of the figure show that decreasing $v_{T}$ does not yield a large disagreement between the final DM central densities. This is because both models experience similar rates of mass loss, despite the difference in their orbits. Differently, the model with $50\%$ larger $v_{T}$, orbits around the MW only once and it does not lose as much mass from tidal interactions, therefore the gravothermal collapse is delayed, reaching an order of magnitude lower $\rho_{150}$ at present time, with respect to the default model.

We conclude that the uncertainties in the galaxies tangential velocities can change the range of cross sections that reproduce the central DM densities, altering the cross section-velocity dependence. In a coming study we will analyse the evolution of truncated SIDM density profiles from gravitational tidal stripping and ram pressure, to further improve the modelling of subhaloes hosting the local dwarf galaxies and provide robust constraints of $\sigma/m_{\chi}$ on dwarf galaxy scales. Such constraints will be adjusted by the uncertainties of the orbital parameters.

\section{Conclusions}\label{Conclusions}

Self-Interacting Dark Matter (SIDM) offers a promising solution to the small-scale challenges faced by the otherwise-remarkably successful Cold Dark Matter (CDM) model. However, robust constraints of the SIDM scattering cross section per unit mass, $\sigma/m_{\chi}$, on dwarf galaxy scales seem to be missing. The anti-correlation between the central DM densities of the bright Milky Way (MW) dwarf spheroidal galaxies (dSph) and their orbital pericenter distances (\citealt{Kaplinghat19}), poses a potential signature of SIDM. In this work, we have investigated such possibility and found that there is a cross section-velocity relation that is able to explain the diverse DM profiles of MW dSph satellites, and is consistent with observational constraints on larger scales.

To model the evolution of SIDM subhaloes hosting the MW dSphs, we have applied the gravothermal fluid formalism for isolated, spherically symmetric self-gravitating SIDM haloes, and extended it to include the orbital evolution around the MW gravitational potential, along with a consistent characterization of gravitational tidal stripping. We have adopted the proper motions from the Gaia mission and used the code galpy (\citealt{Bovy15}) to integrate the orbits of the dSphs. We have also used the fitting functions from \citet{Green19} to model the truncation of subhaloes' density from tidal mass loss. Our model has the advantage of tracking the subhalo evolution within scales smaller than 100 pc, largely expensive to resolve with N-body simulations, while easily covering a wide range of parameter space for the SIDM scattering cross section per unit mass, $\sigma/m_{\chi}$.

We have applied the model to the classical dSph galaxies, namely, Ursa Minor, Draco, Sculptor, Sextans. Fornax, Carina, LeoII, LeoI and Canes Venatici I, using the orbital parameters estimated by \citet{Fritz18} (Fig.\ref{OrbitsDwarfGalaxies}). The subhaloes hosting the dSphs are modelled from an initial NFW profile, virial mass of $(0.1-4)\times 10^{9}\Msun$ and concentration parameter of $6-7$ (Table~\ref{Initial_conditions_table}). Their evolution begins when the Universe is 3.5 Gyr old (redshift $z=1.87$), at a point when none the dSphs have crossed the MW's virial radius, and continues for 10.2 Gyrs until present time.

Along with the virial mass and concentration, subhaloes are initialised with a constant $\sigma/m_{\chi}$. We have shown that the evolution of the subhaloes density profile, as well as the rate of mass loss, largely depends on $\sigma/m_{\chi}$, so that a large $\sigma/m_{\chi}$ leads to a larger rate of DM-DM collisions that produce a shallower and lower density core. We have exemplified the core expansion phase and gravothermal collapse phase with the subhalo hosting a Carina-like dSph (Fig.~\ref{Result_evol_profile}), and also showed the dependence of the central DM density of Carina with $\sigma/m_{\chi}$ (Fig.~\ref{Result_central_density_Carina}). As expected, not only for Carina, but for the remaining dSphs, mass loss from tidal interactions leads to a faster gravothermal core collapse (Fig.~\ref{Result_central_density_all}). 

We have investigated the range of $\sigma/m_{\chi}$ that produces subhaloes with central DM densities in agreement with the observational estimations. There is not single range of $\sigma/m_{\chi}$ able to reproduce the observed data, instead each subhalo is characterized by a specific range. Draco, for instance, prefers lower values of $\sigma/m_{\chi}$, ranging between $\sigma/m_{\chi}$ = 23 cm$^{2}$g$^{-1}$ and $\sigma/m_{\chi}$ = 25 cm$^{2}$g$^{-1}$, whereas LeoII requires $\sigma/m_{\chi}$ ranging between $\sigma/m_{\chi}$ = 120 cm$^{2}$g$^{-1}$ and $\sigma/m_{\chi}$= 90 cm$^{2}$g$^{-1}$. 

Interestingly, the range of $\sigma/m_{\chi}$ needed to reproduce the observed DM densities is not random, instead it correlates with the average collision velocity of DM particles within each subhalo's core. This result points towards an intrinsic velocity-dependent relation (Fig.~\ref{Velocity_cross_section_plane}), that can be fitted by a particle physics model for SIDM, where DM particles of mass $m_{\chi}$ interact under the exchange of a light mediator $\phi$, with the self-interactions being described by a Yukawa potential. By assuming $\alpha_{\chi}=0.01$ (analog of the fine-structure constant in the dark sector), the relation is best fitted by a DM particle mass of $m_{\chi} = 0.648 \pm 0.154$ GeV and a mediator mass of $m_{\phi}= 0.636 \pm 0.055$ MeV. We show that the $\sigma/m_{\chi}$-velocity relation is a feasible velocity-dependent model for SIDM that lies in perfect agreement with current observational constraints on larger scales (Fig.~\ref{Velocity_cross_section_plane_ALL}).

We have extensively analysed the caveats of the model, such as the uncertainty of the free-parameters $b$ and $C$ from the gravothermal modelling, the density truncation time, the MW mass, errors in the orbital parameters, impact of initial subhalo mass, $M_{200,\rm{init}}$, and concentration, $c_{200,\rm{init}}$, as well as different initializations of the density profile. We have shown that our key results are robust to different initializations of the density profile (Fig.~\ref{Comparison_Initial_Profile}), that can either be a NFW profile or core profile. Changes of the parameters $C$ and $b$, the density truncation time and MW mass, modify the ranges of $\sigma/m_{\chi}$ that reproduce the observational estimates, by either increasing or decreasing the overall normalization in up to a factor of 1.5 for $C$ and $b$ (Fig.~\ref{Comparison_C_cte}), in up to a factor of 1.13 for the truncation time (Fig.~\ref{Comparison_Truncation_times}), or in up to a factor of 1.2 for the MW mass (Fig.~\ref{Comparison_MW_mass_fig}). These changes, nevertheless, do not alter the shape of the cross section-velocity relation, only its normalization. 

The initial subhalo mass and concentration can also modify the ranges of $\sigma/m_{\chi}$ that reproduce the observations. Models with lower $M_{200,\rm{init}}$ reach lower central DM densities at present time (Fig.~\ref{Comparison_Initial_Mass}), therefore higher cross sections would be needed to reproduce the observations. We have argued, however, that our choices of $M_{200,\rm{init}}$ are not random, they are tuned so that the systems, in their final state, have a virial mass that reproduces the observational estimations. The concentration parameter controls the core-collapse time-scale, which is shorter for low concentration systems. We have shown that initializing the models with higher concentrations require much lower $\sigma/m_{\chi}$ to reproduce the observed central DM densities (Fig.~\ref{Comparison_c200}). However, we believe that such large initial concentrations are not well justified. At high redshift ($z>1$), the halo has large rates of accretion, with its mass history mainly characterized by an exponential growth. During this time, the scale radius increases simultaneously with the virial radius, hence the concentration hardly grows (\citealt{Correa15c}).

Finally, we have addressed the impact in the uncertainty of the orbital parameters estimated by the Gaia Collaboration (\citealt{Fritz18,Brown18,Helmi18}). While our results are robust to the small uncertainties in the galactocentric distances and radial velocities, they are not to the larger uncertainties of the galaxies tangential velocities (Fig.~\ref{Impact_vT_fig}), that can potentially weaken the cross-section velocity correlation, as well as the anti-correlation of central DM densities with pericenter passage reported by \citet{Kaplinghat19}. 

In this paper we have made a first assessment of the viability of a velocity-dependent SIDM model able to explain a specific observable, the anti-correlation between the central DM densities of the bright MW dSph and their orbital pericenter distances. We have found that there is such model, that explains the diverse DM profiles of MW dSph satellites, is consistent with observational constraints on larger scales and predicts that the dSphs are undergoing gravothermal collapse. However more evidence will be gathered in a coming study, to further support or exclude such scenario. We will also assess the impact of baryons, as well as the evolution of truncated SIDM density profiles from gravitational tidal stripping and ram pressure, to further improve the modelling of subhaloes hosting the local dwarf galaxies and provide robust constraints of $\sigma/m_{\chi}$ on dwarf galaxy scales. Such constraints will be adjusted by the uncertainties in the orbital parameters.

\section*{Acknowledgements}

It is a pleasure to thank David Harvey for a careful reading of the manuscript and enlightening comments. This work was supported by the Dutch Research Council (NWO Veni 192.020). I acknowledge various public python packages that have greatly benefited this work: \verb|scipy| (Jones et al. 2001), \verb|numpy| (van der Walt et al. 2011), \verb|matplotlib| (Hunter 2007) and \verb|ipython| (Perez \& Granger 2007). 

\section*{Data availability}

The data supporting the plots within this article are available on reasonable request to the corresponding author.

\bibliography{biblio}

\begin{thebibliography}{}
\makeatletter
\relax
\def\mn@urlcharsother{\let\do\@makeother \do\$\do\&\do\#\do\^\do\_\do\%\do\~}
\def\mn@doi{\begingroup\mn@urlcharsother \@ifnextchar [ {\mn@doi@}
  {\mn@doi@[]}}
\def\mn@doi@[#1]#2{\def\@tempa{#1}\ifx\@tempa\@empty \href
  {http://dx.doi.org/#2} {doi:#2}\else \href {http://dx.doi.org/#2} {#1}\fi
  \endgroup}
\def\mn@eprint#1#2{\mn@eprint@#1:#2::\@nil}
\def\mn@eprint@arXiv#1{\href {http://arxiv.org/abs/#1} {{\tt arXiv:#1}}}
\def\mn@eprint@dblp#1{\href {http://dblp.uni-trier.de/rec/bibtex/#1.xml}
  {dblp:#1}}
\def\mn@eprint@#1:#2:#3:#4\@nil{\def\@tempa {#1}\def\@tempb {#2}\def\@tempc
  {#3}\ifx \@tempc \@empty \let \@tempc \@tempb \let \@tempb \@tempa \fi \ifx
  \@tempb \@empty \def\@tempb {arXiv}\fi \@ifundefined
  {mn@eprint@\@tempb}{\@tempb:\@tempc}{\expandafter \expandafter \csname
  mn@eprint@\@tempb\endcsname \expandafter{\@tempc}}}

\bibitem[\protect\citeauthoryear{{Arkani-Hamed}, {Finkbeiner}, {Slatyer}  \&
  {Weiner}}{{Arkani-Hamed} et~al.}{2009}]{ArkaniHamed09}
{Arkani-Hamed} N.,  {Finkbeiner} D.~P.,  {Slatyer} T.~R.,   {Weiner} N.,  2009,
  \mn@doi [\prd] {10.1103/PhysRevD.79.015014}, \href
  {https://ui.adsabs.harvard.edu/abs/2009PhRvD..79a5014A} {79, 015014}

\bibitem[\protect\citeauthoryear{{Balberg}, {Shapiro}  \& {Inagaki}}{{Balberg}
  et~al.}{2002}]{Balberg02}
{Balberg} S.,  {Shapiro} S.~L.,   {Inagaki} S.,  2002, \mn@doi [\apj]
  {10.1086/339038}, \href
  {https://ui.adsabs.harvard.edu/abs/2002ApJ...568..475B} {568, 475}

\bibitem[\protect\citeauthoryear{{Banerjee}, {Adhikari}, {Dalal}, {More}  \&
  {Kravtsov}}{{Banerjee} et~al.}{2020}]{Banerjee20}
{Banerjee} A.,  {Adhikari} S.,  {Dalal} N.,  {More} S.,   {Kravtsov} A.,  2020,
  \mn@doi [\jcap] {10.1088/1475-7516/2020/02/024}, \href
  {https://ui.adsabs.harvard.edu/abs/2020JCAP...02..024B} {2020, 024}

\bibitem[\protect\citeauthoryear{{Boddy}, {Feng}, {Kaplinghat}, {Shadmi}  \&
  {Tait}}{{Boddy} et~al.}{2014}]{Boddy14}
{Boddy} K.~K.,  {Feng} J.~L.,  {Kaplinghat} M.,  {Shadmi} Y.,   {Tait} T.
  M.~P.,  2014, \mn@doi [\prd] {10.1103/PhysRevD.90.095016}, \href
  {https://ui.adsabs.harvard.edu/abs/2014PhRvD..90i5016B} {90, 095016}

\bibitem[\protect\citeauthoryear{{Bovy}}{{Bovy}}{2015}]{Bovy15}
{Bovy} J.,  2015, \mn@doi [\apjs] {10.1088/0067-0049/216/2/29}, \href
  {https://ui.adsabs.harvard.edu/abs/2015ApJS..216...29B} {216, 29}

\bibitem[\protect\citeauthoryear{{Boylan-Kolchin}, {Bullock}  \&
  {Kaplinghat}}{{Boylan-Kolchin} et~al.}{2011}]{BoylanKolchin11}
{Boylan-Kolchin} M.,  {Bullock} J.~S.,   {Kaplinghat} M.,  2011, \mn@doi
  [\mnras] {10.1111/j.1745-3933.2011.01074.x}, \href
  {https://ui.adsabs.harvard.edu/abs/2011MNRAS.415L..40B} {415, L40}

\bibitem[\protect\citeauthoryear{{Boylan-Kolchin}, {Bullock}  \&
  {Kaplinghat}}{{Boylan-Kolchin} et~al.}{2012}]{BoylanKolchin12}
{Boylan-Kolchin} M.,  {Bullock} J.~S.,   {Kaplinghat} M.,  2012, \mn@doi
  [\mnras] {10.1111/j.1365-2966.2012.20695.x}, \href
  {https://ui.adsabs.harvard.edu/abs/2012MNRAS.422.1203B} {422, 1203}

\bibitem[\protect\citeauthoryear{{Brown} et~al.,}{{Brown}
  et~al.}{2018}]{Brown18}
{Brown} A.~G.~A.,  et~al., 2018, \mn@doi [\aap] {10.1051/0004-6361/201833051},
  \href {https://ui.adsabs.harvard.edu/abs/2018A&A...616A...1G} {616, A1}

\bibitem[\protect\citeauthoryear{{Buckley} \& {Fox}}{{Buckley} \&
  {Fox}}{2010}]{Buckley10}
{Buckley} M.~R.,  {Fox} P.~J.,  2010, \mn@doi [\prd]
  {10.1103/PhysRevD.81.083522}, \href
  {https://ui.adsabs.harvard.edu/abs/2010PhRvD..81h3522B} {81, 083522}

\bibitem[\protect\citeauthoryear{{Buckley}, {Zavala}, {Cyr-Racine}, {Sigurdson}
   \& {Vogelsberger}}{{Buckley} et~al.}{2014}]{Buckley14}
{Buckley} M.~R.,  {Zavala} J.,  {Cyr-Racine} F.-Y.,  {Sigurdson} K.,
  {Vogelsberger} M.,  2014, \mn@doi [\prd] {10.1103/PhysRevD.90.043524}, \href
  {https://ui.adsabs.harvard.edu/abs/2014PhRvD..90d3524B} {90, 043524}

\bibitem[\protect\citeauthoryear{{Chapman}, {Cowling}, {Burnett}  \&
  {Cercignani}}{{Chapman} et~al.}{1990}]{Chapman90}
{Chapman} S.,  {Cowling} T.,  {Burnett} D.,   {Cercignani} C.,  1990, {The
  Mathematical Theory of Non-uniform Gases: An Account of the Kinetic Theory of
  Viscosity, Thermal Conduction and Diffusion in Gases}.
Cambridge Mathematical Library, Cambridge University Press, \url
  {https://books.google.nl/books?id=Cbp5JP2OTrwC}

\bibitem[\protect\citeauthoryear{{Col{\'\i}n}, {Avila-Reese}, {Valenzuela}  \&
  {Firmani}}{{Col{\'\i}n} et~al.}{2002}]{Colin02}
{Col{\'\i}n} P.,  {Avila-Reese} V.,  {Valenzuela} O.,   {Firmani} C.,  2002,
  \mn@doi [\apj] {10.1086/344259}, \href
  {https://ui.adsabs.harvard.edu/abs/2002ApJ...581..777C} {581, 777}

\bibitem[\protect\citeauthoryear{{Correa}, {Wyithe}, {Schaye}  \&
  {Duffy}}{{Correa} et~al.}{2015a}]{Correa15a}
{Correa} C.~A.,  {Wyithe} J. S.~B.,  {Schaye} J.,   {Duffy} A.~R.,  2015a,
  \mn@doi [\mnras] {10.1093/mnras/stv689}, \href
  {https://ui.adsabs.harvard.edu/abs/2015MNRAS.450.1514C} {450, 1514}

\bibitem[\protect\citeauthoryear{{Correa}, {Wyithe}, {Schaye}  \&
  {Duffy}}{{Correa} et~al.}{2015b}]{Correa15c}
{Correa} C.~A.,  {Wyithe} J. S.~B.,  {Schaye} J.,   {Duffy} A.~R.,  2015b,
  \mn@doi [\mnras] {10.1093/mnras/stv697}, \href
  {https://ui.adsabs.harvard.edu/abs/2015MNRAS.450.1521C} {450, 1521}

\bibitem[\protect\citeauthoryear{{Correa}, {Wyithe}, {Schaye}  \&
  {Duffy}}{{Correa} et~al.}{2015c}]{Correa15b}
{Correa} C.~A.,  {Wyithe} J. S.~B.,  {Schaye} J.,   {Duffy} A.~R.,  2015c,
  \mn@doi [\mnras] {10.1093/mnras/stv1363}, \href
  {https://ui.adsabs.harvard.edu/abs/2015MNRAS.452.1217C} {452, 1217}

\bibitem[\protect\citeauthoryear{{Cyr-Racine}, {Sigurdson}, {Zavala},
  {Bringmann}, {Vogelsberger}  \& {Pfrommer}}{{Cyr-Racine}
  et~al.}{2016}]{CyrRacine16}
{Cyr-Racine} F.-Y.,  {Sigurdson} K.,  {Zavala} J.,  {Bringmann} T.,
  {Vogelsberger} M.,   {Pfrommer} C.,  2016, \mn@doi [\prd]
  {10.1103/PhysRevD.93.123527}, \href
  {https://ui.adsabs.harvard.edu/abs/2016PhRvD..93l3527C} {93, 123527}

\bibitem[\protect\citeauthoryear{{Dav{\'e}}, {Spergel}, {Steinhardt}  \&
  {Wandelt}}{{Dav{\'e}} et~al.}{2001}]{Dave01}
{Dav{\'e}} R.,  {Spergel} D.~N.,  {Steinhardt} P.~J.,   {Wandelt} B.~D.,  2001,
  \mn@doi [\apj] {10.1086/318417}, \href
  {https://ui.adsabs.harvard.edu/abs/2001ApJ...547..574D} {547, 574}

\bibitem[\protect\citeauthoryear{{Dooley}, {Peter}, {Vogelsberger}, {Zavala}
  \& {Frebel}}{{Dooley} et~al.}{2016}]{Dooley16}
{Dooley} G.~A.,  {Peter} A. H.~G.,  {Vogelsberger} M.,  {Zavala} J.,   {Frebel}
  A.,  2016, \mn@doi [\mnras] {10.1093/mnras/stw1309}, \href
  {https://ui.adsabs.harvard.edu/abs/2016MNRAS.461..710D} {461, 710}

\bibitem[\protect\citeauthoryear{{Dutton}, {Macci{\`o}}, {Frings}, {Wang},
  {Stinson}, {Penzo}  \& {Kang}}{{Dutton} et~al.}{2016}]{Dutton16}
{Dutton} A.~A.,  {Macci{\`o}} A.~V.,  {Frings} J.,  {Wang} L.,  {Stinson}
  G.~S.,  {Penzo} C.,   {Kang} X.,  2016, \mn@doi [\mnras]
  {10.1093/mnrasl/slv193}, \href
  {https://ui.adsabs.harvard.edu/abs/2016MNRAS.457L..74D} {457, L74}

\bibitem[\protect\citeauthoryear{{Elbert}, {Bullock}, {Garrison-Kimmel},
  {Rocha}, {O{\~n}orbe}  \& {Peter}}{{Elbert} et~al.}{2015}]{Elbert15}
{Elbert} O.~D.,  {Bullock} J.~S.,  {Garrison-Kimmel} S.,  {Rocha} M.,
  {O{\~n}orbe} J.,   {Peter} A. H.~G.,  2015, \mn@doi [\mnras]
  {10.1093/mnras/stv1470}, \href
  {https://ui.adsabs.harvard.edu/abs/2015MNRAS.453...29E} {453, 29}

\bibitem[\protect\citeauthoryear{{Essig}, {McDermott}, {Yu}  \&
  {Zhong}}{{Essig} et~al.}{2019}]{Essig19}
{Essig} R.,  {McDermott} S.~D.,  {Yu} H.-B.,   {Zhong} Y.-M.,  2019, \mn@doi
  [\prl] {10.1103/PhysRevLett.123.121102}, \href
  {https://ui.adsabs.harvard.edu/abs/2019PhRvL.123l1102E} {123, 121102}

\bibitem[\protect\citeauthoryear{{Fattahi} et~al.,}{{Fattahi}
  et~al.}{2016}]{Fattahi16}
{Fattahi} A.,  et~al., 2016, \mn@doi [\mnras] {10.1093/mnras/stv2970}, \href
  {https://ui.adsabs.harvard.edu/abs/2016MNRAS.457..844F} {457, 844}

\bibitem[\protect\citeauthoryear{{Feng}, {Kaplinghat}  \& {Yu}}{{Feng}
  et~al.}{2010}]{Feng10}
{Feng} J.~L.,  {Kaplinghat} M.,   {Yu} H.-B.,  2010, \mn@doi [\prd]
  {10.1103/PhysRevD.82.083525}, \href
  {https://ui.adsabs.harvard.edu/abs/2010PhRvD..82h3525F} {82, 083525}

\bibitem[\protect\citeauthoryear{{Flores} \& {Primack}}{{Flores} \&
  {Primack}}{1994}]{Flores94}
{Flores} R.~A.,  {Primack} J.~R.,  1994, \mn@doi [\apjl] {10.1086/187350},
  \href {https://ui.adsabs.harvard.edu/abs/1994ApJ...427L...1F} {427, L1}

\bibitem[\protect\citeauthoryear{{Fritz}, {Battaglia}, {Pawlowski},
  {Kallivayalil}, {van der Marel}, {Sohn}, {Brook}  \& {Besla}}{{Fritz}
  et~al.}{2018}]{Fritz18}
{Fritz} T.~K.,  {Battaglia} G.,  {Pawlowski} M.~S.,  {Kallivayalil} N.,  {van
  der Marel} R.,  {Sohn} S.~T.,  {Brook} C.,   {Besla} G.,  2018, \mn@doi
  [\aap] {10.1051/0004-6361/201833343}, \href
  {https://ui.adsabs.harvard.edu/abs/2018A&A...619A.103F} {619, A103}

\bibitem[\protect\citeauthoryear{{Garrison-Kimmel}, {Boylan-Kolchin}, {Bullock}
   \& {Kirby}}{{Garrison-Kimmel} et~al.}{2014}]{GarrisonKimmel14}
{Garrison-Kimmel} S.,  {Boylan-Kolchin} M.,  {Bullock} J.~S.,   {Kirby} E.~N.,
  2014, \mn@doi [\mnras] {10.1093/mnras/stu1477}, \href
  {https://ui.adsabs.harvard.edu/abs/2014MNRAS.444..222G} {444, 222}

\bibitem[\protect\citeauthoryear{{Green} \& {van den Bosch}}{{Green} \& {van
  den Bosch}}{2019}]{Green19}
{Green} S.~B.,  {van den Bosch} F.~C.,  2019, \mn@doi [\mnras]
  {10.1093/mnras/stz2767}, \href
  {https://ui.adsabs.harvard.edu/abs/2019MNRAS.490.2091G} {490, 2091}

\bibitem[\protect\citeauthoryear{{Han}, {Cole}, {Frenk}  \& {Jing}}{{Han}
  et~al.}{2016}]{Han16}
{Han} J.,  {Cole} S.,  {Frenk} C.~S.,   {Jing} Y.,  2016, \mn@doi [\mnras]
  {10.1093/mnras/stv2900}, \href
  {https://ui.adsabs.harvard.edu/abs/2016MNRAS.457.1208H} {457, 1208}

\bibitem[\protect\citeauthoryear{{Harvey}, {Massey}, {Kitching}, {Taylor}  \&
  {Tittley}}{{Harvey} et~al.}{2015}]{Harvey15}
{Harvey} D.,  {Massey} R.,  {Kitching} T.,  {Taylor} A.,   {Tittley} E.,  2015,
  \mn@doi [Science] {10.1126/science.1261381}, \href
  {https://ui.adsabs.harvard.edu/abs/2015Sci...347.1462H} {347, 1462}

\bibitem[\protect\citeauthoryear{{Harvey}, {Revaz}, {Robertson}  \&
  {Hausammann}}{{Harvey} et~al.}{2018}]{Harvey18}
{Harvey} D.,  {Revaz} Y.,  {Robertson} A.,   {Hausammann} L.,  2018, \mn@doi
  [\mnras] {10.1093/mnrasl/sly159}, \href
  {https://ui.adsabs.harvard.edu/abs/2018MNRAS.481L..89H} {481, L89}

\bibitem[\protect\citeauthoryear{{Harvey}, {Robertson}, {Massey}  \&
  {McCarthy}}{{Harvey} et~al.}{2019}]{Harvey19}
{Harvey} D.,  {Robertson} A.,  {Massey} R.,   {McCarthy} I.~G.,  2019, \mn@doi
  [\mnras] {10.1093/mnras/stz1816}, \href
  {https://ui.adsabs.harvard.edu/abs/2019MNRAS.488.1572H} {488, 1572}

\bibitem[\protect\citeauthoryear{{Helmi} et~al.,}{{Helmi}
  et~al.}{2018}]{Helmi18}
{Helmi} A.,  et~al., 2018, \mn@doi [\aap] {10.1051/0004-6361/201832698}, \href
  {https://ui.adsabs.harvard.edu/abs/2018A&A...616A..12G} {616, A12}

\bibitem[\protect\citeauthoryear{{Ibe} \& {Yu}}{{Ibe} \& {Yu}}{2010}]{Ibe10}
{Ibe} M.,  {Yu} H.-B.,  2010, \mn@doi [Physics Letters B]
  {10.1016/j.physletb.2010.07.026}, \href
  {https://ui.adsabs.harvard.edu/abs/2010PhLB..692...70I} {692, 70}

\bibitem[\protect\citeauthoryear{{Jardel} \& {Gebhardt}}{{Jardel} \&
  {Gebhardt}}{2012}]{Jardel12}
{Jardel} J.~R.,  {Gebhardt} K.,  2012, \mn@doi [\apj]
  {10.1088/0004-637X/746/1/89}, \href
  {https://ui.adsabs.harvard.edu/abs/2012ApJ...746...89J} {746, 89}

\bibitem[\protect\citeauthoryear{{Jiang} \& {van den Bosch}}{{Jiang} \& {van
  den Bosch}}{2017}]{Jiang17}
{Jiang} F.,  {van den Bosch} F.~C.,  2017, \mn@doi [\mnras]
  {10.1093/mnras/stx1979}, \href
  {https://ui.adsabs.harvard.edu/abs/2017MNRAS.472..657J} {472, 657}

\bibitem[\protect\citeauthoryear{{Kahlhoefer}, {Schmidt-Hoberg}, {Kummer}  \&
  {Sarkar}}{{Kahlhoefer} et~al.}{2015}]{Kahlhoefer15}
{Kahlhoefer} F.,  {Schmidt-Hoberg} K.,  {Kummer} J.,   {Sarkar} S.,  2015,
  \mn@doi [\mnras] {10.1093/mnrasl/slv088}, \href
  {https://ui.adsabs.harvard.edu/abs/2015MNRAS.452L..54K} {452, L54}

\bibitem[\protect\citeauthoryear{{Kahlhoefer}, {Kaplinghat}, {Slatyer}  \&
  {Wu}}{{Kahlhoefer} et~al.}{2019}]{Kahlhoefer19}
{Kahlhoefer} F.,  {Kaplinghat} M.,  {Slatyer} T.~R.,   {Wu} C.-L.,  2019,
  \mn@doi [\jcap] {10.1088/1475-7516/2019/12/010}, \href
  {https://ui.adsabs.harvard.edu/abs/2019JCAP...12..010K} {2019, 010}

\bibitem[\protect\citeauthoryear{{Kamada}, {Kaplinghat}, {Pace}  \&
  {Yu}}{{Kamada} et~al.}{2017}]{Kamada17}
{Kamada} A.,  {Kaplinghat} M.,  {Pace} A.~B.,   {Yu} H.-B.,  2017, \mn@doi
  [\prl] {10.1103/PhysRevLett.119.111102}, \href
  {https://ui.adsabs.harvard.edu/abs/2017PhRvL.119k1102K} {119, 111102}

\bibitem[\protect\citeauthoryear{{Kaplinghat}, {Tulin}  \& {Yu}}{{Kaplinghat}
  et~al.}{2016}]{Kaplinghat16}
{Kaplinghat} M.,  {Tulin} S.,   {Yu} H.-B.,  2016, \mn@doi [\prl]
  {10.1103/PhysRevLett.116.041302}, \href
  {https://ui.adsabs.harvard.edu/abs/2016PhRvL.116d1302K} {116, 041302}

\bibitem[\protect\citeauthoryear{{Kaplinghat}, {Valli}  \& {Yu}}{{Kaplinghat}
  et~al.}{2019}]{Kaplinghat19}
{Kaplinghat} M.,  {Valli} M.,   {Yu} H.-B.,  2019, \mn@doi [\mnras]
  {10.1093/mnras/stz2511}, \href
  {https://ui.adsabs.harvard.edu/abs/2019MNRAS.490..231K} {490, 231}

\bibitem[\protect\citeauthoryear{{Kim}, {Peter}  \& {Wittman}}{{Kim}
  et~al.}{2017}]{Kim17}
{Kim} S.~Y.,  {Peter} A. H.~G.,   {Wittman} D.,  2017, \mn@doi [\mnras]
  {10.1093/mnras/stx896}, \href
  {https://ui.adsabs.harvard.edu/abs/2017MNRAS.469.1414K} {469, 1414}

\bibitem[\protect\citeauthoryear{{King}}{{King}}{1962}]{King62}
{King} I.,  1962, \mn@doi [\aj] {10.1086/108756}, \href
  {https://ui.adsabs.harvard.edu/abs/1962AJ.....67..471K} {67, 471}

\bibitem[\protect\citeauthoryear{{Klypin}, {Kravtsov}, {Valenzuela}  \&
  {Prada}}{{Klypin} et~al.}{1999}]{Klypin99}
{Klypin} A.,  {Kravtsov} A.~V.,  {Valenzuela} O.,   {Prada} F.,  1999, \mn@doi
  [\apj] {10.1086/307643}, \href
  {https://ui.adsabs.harvard.edu/abs/1999ApJ...522...82K} {522, 82}

\bibitem[\protect\citeauthoryear{{Koda} \& {Shapiro}}{{Koda} \&
  {Shapiro}}{2011}]{Koda11}
{Koda} J.,  {Shapiro} P.~R.,  2011, \mn@doi [\mnras]
  {10.1111/j.1365-2966.2011.18684.x}, \href
  {https://ui.adsabs.harvard.edu/abs/2011MNRAS.415.1125K} {415, 1125}

\bibitem[\protect\citeauthoryear{{Lynden-Bell} \& {Eggleton}}{{Lynden-Bell} \&
  {Eggleton}}{1980}]{LyndenBell80}
{Lynden-Bell} D.,  {Eggleton} P.~P.,  1980, \mn@doi [\mnras]
  {10.1093/mnras/191.3.483}, \href
  {https://ui.adsabs.harvard.edu/abs/1980MNRAS.191..483L} {191, 483}

\bibitem[\protect\citeauthoryear{{Miralda-Escud{\'e}}}{{Miralda-Escud{\'e}}}{2002}]{MiraldaEscude02}
{Miralda-Escud{\'e}} J.,  2002, \mn@doi [\apj] {10.1086/324138}, \href
  {https://ui.adsabs.harvard.edu/abs/2002ApJ...564...60M} {564, 60}

\bibitem[\protect\citeauthoryear{{Moore}}{{Moore}}{1994}]{Moore94}
{Moore} B.,  1994, \mn@doi [\nat] {10.1038/370629a0}, \href
  {https://ui.adsabs.harvard.edu/abs/1994Natur.370..629M} {370, 629}

\bibitem[\protect\citeauthoryear{{Moore}, {Quinn}, {Governato}, {Stadel}  \&
  {Lake}}{{Moore} et~al.}{1999}]{Moore99}
{Moore} B.,  {Quinn} T.,  {Governato} F.,  {Stadel} J.,   {Lake} G.,  1999,
  \mn@doi [\mnras] {10.1046/j.1365-8711.1999.03039.x}, \href
  {https://ui.adsabs.harvard.edu/abs/1999MNRAS.310.1147M} {310, 1147}

\bibitem[\protect\citeauthoryear{{Nadler}, {Banerjee}, {Adhikari}, {Mao}  \&
  {Wechsler}}{{Nadler} et~al.}{2020}]{Nadler20}
{Nadler} E.~O.,  {Banerjee} A.,  {Adhikari} S.,  {Mao} Y.-Y.,   {Wechsler}
  R.~H.,  2020, arXiv e-prints, \href
  {https://ui.adsabs.harvard.edu/abs/2020arXiv200108754N} {p. arXiv:2001.08754}

\bibitem[\protect\citeauthoryear{{Navarro}, {Eke}  \& {Frenk}}{{Navarro}
  et~al.}{1996}]{Navarro96}
{Navarro} J.~F.,  {Eke} V.~R.,   {Frenk} C.~S.,  1996, \mn@doi [\mnras]
  {10.1093/mnras/283.3.L72}, \href
  {https://ui.adsabs.harvard.edu/abs/1996MNRAS.283L..72N} {283, L72}

\bibitem[\protect\citeauthoryear{{Navarro}, {Frenk}  \& {White}}{{Navarro}
  et~al.}{1997}]{NFW97}
{Navarro} J.~F.,  {Frenk} C.~S.,   {White} S. D.~M.,  1997, \mn@doi [\apj]
  {10.1086/304888}, \href
  {https://ui.adsabs.harvard.edu/abs/1997ApJ...490..493N} {490, 493}

\bibitem[\protect\citeauthoryear{{Nishikawa}, {Boddy}  \&
  {Kaplinghat}}{{Nishikawa} et~al.}{2020}]{Nishikawa19}
{Nishikawa} H.,  {Boddy} K.~K.,   {Kaplinghat} M.,  2020, \mn@doi [\prd]
  {10.1103/PhysRevD.101.063009}, \href
  {https://ui.adsabs.harvard.edu/abs/2020PhRvD.101f3009N} {101, 063009}

\bibitem[\protect\citeauthoryear{{Ogiya}, {van den Bosch}, {Hahn}, {Green},
  {Miller}  \& {Burkert}}{{Ogiya} et~al.}{2019}]{Ogiya19}
{Ogiya} G.,  {van den Bosch} F.~C.,  {Hahn} O.,  {Green} S.~B.,  {Miller}
  T.~B.,   {Burkert} A.,  2019, \mn@doi [\mnras] {10.1093/mnras/stz375}, \href
  {https://ui.adsabs.harvard.edu/abs/2019MNRAS.485..189O} {485, 189}

\bibitem[\protect\citeauthoryear{{Papastergis}, {Giovanelli}, {Haynes}  \&
  {Shankar}}{{Papastergis} et~al.}{2015}]{Papastergis15}
{Papastergis} E.,  {Giovanelli} R.,  {Haynes} M.~P.,   {Shankar} F.,  2015,
  \mn@doi [\aap] {10.1051/0004-6361/201424909}, \href
  {https://ui.adsabs.harvard.edu/abs/2015A&A...574A.113P} {574, A113}

\bibitem[\protect\citeauthoryear{{Pascale}, {Posti}, {Nipoti}  \&
  {Binney}}{{Pascale} et~al.}{2018}]{Pascale18}
{Pascale} R.,  {Posti} L.,  {Nipoti} C.,   {Binney} J.,  2018, \mn@doi [\mnras]
  {10.1093/mnras/sty1860}, \href
  {https://ui.adsabs.harvard.edu/abs/2018MNRAS.480..927P} {480, 927}

\bibitem[\protect\citeauthoryear{{Peter}, {Rocha}, {Bullock}  \&
  {Kaplinghat}}{{Peter} et~al.}{2013}]{Peter13}
{Peter} A. H.~G.,  {Rocha} M.,  {Bullock} J.~S.,   {Kaplinghat} M.,  2013,
  \mn@doi [\mnras] {10.1093/mnras/sts535}, \href
  {https://ui.adsabs.harvard.edu/abs/2013MNRAS.430..105P} {430, 105}

\bibitem[\protect\citeauthoryear{{Planck Collaboration} et~al.}{{Planck
  Collaboration} et~al.}{2014}]{Planck}
{Planck Collaboration} et~al., 2014, \mn@doi [\aap]
  {10.1051/0004-6361/201321591}, \href
  {http://adsabs.harvard.edu/abs/2014A%26A...571A..16P} {571, A16}

\bibitem[\protect\citeauthoryear{{Pontzen} \& {Governato}}{{Pontzen} \&
  {Governato}}{2012}]{Pontzen12}
{Pontzen} A.,  {Governato} F.,  2012, \mn@doi [\mnras]
  {10.1111/j.1365-2966.2012.20571.x}, \href
  {https://ui.adsabs.harvard.edu/abs/2012MNRAS.421.3464P} {421, 3464}

\bibitem[\protect\citeauthoryear{{Pospelov}, {Ritz}  \& {Voloshin}}{{Pospelov}
  et~al.}{2008}]{Pospelov08}
{Pospelov} M.,  {Ritz} A.,   {Voloshin} M.,  2008, \mn@doi [\prd]
  {10.1103/PhysRevD.78.115012}, \href
  {https://ui.adsabs.harvard.edu/abs/2008PhRvD..78k5012P} {78, 115012}

\bibitem[\protect\citeauthoryear{{Press} \& {Schechter}}{{Press} \&
  {Schechter}}{1974}]{Press74}
{Press} W.~H.,  {Schechter} P.,  1974, \mn@doi [\apj] {10.1086/152650}, \href
  {https://ui.adsabs.harvard.edu/abs/1974ApJ...187..425P} {187, 425}

\bibitem[\protect\citeauthoryear{{Randall}, {Markevitch}, {Clowe}, {Gonzalez}
  \& {Brada{\v{c}}}}{{Randall} et~al.}{2008}]{Randall08}
{Randall} S.~W.,  {Markevitch} M.,  {Clowe} D.,  {Gonzalez} A.~H.,
  {Brada{\v{c}}} M.,  2008, \mn@doi [\apj] {10.1086/587859}, \href
  {https://ui.adsabs.harvard.edu/abs/2008ApJ...679.1173R} {679, 1173}

\bibitem[\protect\citeauthoryear{{Read} \& {Gilmore}}{{Read} \&
  {Gilmore}}{2005}]{Read05}
{Read} J.~I.,  {Gilmore} G.,  2005, \mn@doi [\mnras]
  {10.1111/j.1365-2966.2004.08424.x}, \href
  {https://ui.adsabs.harvard.edu/abs/2005MNRAS.356..107R} {356, 107}

\bibitem[\protect\citeauthoryear{{Read}, {Walker}  \& {Steger}}{{Read}
  et~al.}{2018}]{Read18}
{Read} J.~I.,  {Walker} M.~G.,   {Steger} P.,  2018, \mn@doi [\mnras]
  {10.1093/mnras/sty2286}, \href
  {https://ui.adsabs.harvard.edu/abs/2018MNRAS.481..860R} {481, 860}

\bibitem[\protect\citeauthoryear{{Read}, {Walker}  \& {Steger}}{{Read}
  et~al.}{2019}]{Read19}
{Read} J.~I.,  {Walker} M.~G.,   {Steger} P.,  2019, \mn@doi [\mnras]
  {10.1093/mnras/sty3404}, \href
  {https://ui.adsabs.harvard.edu/abs/2019MNRAS.484.1401R} {484, 1401}

\bibitem[\protect\citeauthoryear{{Ren}, {Kwa}, {Kaplinghat}  \& {Yu}}{{Ren}
  et~al.}{2019}]{Ren19}
{Ren} T.,  {Kwa} A.,  {Kaplinghat} M.,   {Yu} H.-B.,  2019, \mn@doi [Physical
  Review X] {10.1103/PhysRevX.9.031020}, \href
  {https://ui.adsabs.harvard.edu/abs/2019PhRvX...9c1020R} {9, 031020}

\bibitem[\protect\citeauthoryear{{Robertson}, {Massey}  \& {Eke}}{{Robertson}
  et~al.}{2017}]{Robertson17}
{Robertson} A.,  {Massey} R.,   {Eke} V.,  2017, \mn@doi [\mnras]
  {10.1093/mnras/stx463}, \href
  {https://ui.adsabs.harvard.edu/abs/2017MNRAS.467.4719R} {467, 4719}

\bibitem[\protect\citeauthoryear{{Robertson}, {Harvey}, {Massey}, {Eke},
  {McCarthy}, {Jauzac}, {Li}  \& {Schaye}}{{Robertson}
  et~al.}{2019}]{Robertson19}
{Robertson} A.,  {Harvey} D.,  {Massey} R.,  {Eke} V.,  {McCarthy} I.~G.,
  {Jauzac} M.,  {Li} B.,   {Schaye} J.,  2019, \mn@doi [\mnras]
  {10.1093/mnras/stz1815}, \href
  {https://ui.adsabs.harvard.edu/abs/2019MNRAS.488.3646R} {488, 3646}

\bibitem[\protect\citeauthoryear{{Robles}, {Kelley}, {Bullock}  \&
  {Kaplinghat}}{{Robles} et~al.}{2019}]{Robles19}
{Robles} V.~H.,  {Kelley} T.,  {Bullock} J.~S.,   {Kaplinghat} M.,  2019,
  \mn@doi [\mnras] {10.1093/mnras/stz2345}, \href
  {https://ui.adsabs.harvard.edu/abs/2019MNRAS.490.2117R} {490, 2117}

\bibitem[\protect\citeauthoryear{{Rocha}, {Peter}, {Bullock}, {Kaplinghat},
  {Garrison-Kimmel}, {O{\~n}orbe}  \& {Moustakas}}{{Rocha}
  et~al.}{2013}]{Rocha13}
{Rocha} M.,  {Peter} A. H.~G.,  {Bullock} J.~S.,  {Kaplinghat} M.,
  {Garrison-Kimmel} S.,  {O{\~n}orbe} J.,   {Moustakas} L.~A.,  2013, \mn@doi
  [\mnras] {10.1093/mnras/sts514}, \href
  {https://ui.adsabs.harvard.edu/abs/2013MNRAS.430...81R} {430, 81}

\bibitem[\protect\citeauthoryear{{Sameie}, {Yu}, {Sales}, {Vogelsberger}  \&
  {Zavala}}{{Sameie} et~al.}{2020}]{Sameie20}
{Sameie} O.,  {Yu} H.-B.,  {Sales} L.~V.,  {Vogelsberger} M.,   {Zavala} J.,
  2020, \mn@doi [\prl] {10.1103/PhysRevLett.124.141102}, \href
  {https://ui.adsabs.harvard.edu/abs/2020PhRvL.124n1102S} {124, 141102}

\bibitem[\protect\citeauthoryear{{Sawala} et~al.,}{{Sawala}
  et~al.}{2016}]{Sawala16}
{Sawala} T.,  et~al., 2016, \mn@doi [\mnras] {10.1093/mnras/stw145}, \href
  {https://ui.adsabs.harvard.edu/abs/2016MNRAS.457.1931S} {457, 1931}

\bibitem[\protect\citeauthoryear{{Shapiro}}{{Shapiro}}{2018}]{Shapiro18}
{Shapiro} S.~L.,  2018, \mn@doi [\prd] {10.1103/PhysRevD.98.023021}, \href
  {https://ui.adsabs.harvard.edu/abs/2018PhRvD..98b3021S} {98, 023021}

\bibitem[\protect\citeauthoryear{{Spergel} \& {Steinhardt}}{{Spergel} \&
  {Steinhardt}}{2000}]{Spergel00}
{Spergel} D.~N.,  {Steinhardt} P.~J.,  2000, \mn@doi [\prl]
  {10.1103/PhysRevLett.84.3760}, \href
  {https://ui.adsabs.harvard.edu/abs/2000PhRvL..84.3760S} {84, 3760}

\bibitem[\protect\citeauthoryear{{Springel}, {Frenk}  \& {White}}{{Springel}
  et~al.}{2006}]{Springel06}
{Springel} V.,  {Frenk} C.~S.,   {White} S. D.~M.,  2006, \mn@doi [\nat]
  {10.1038/nature04805}, \href
  {https://ui.adsabs.harvard.edu/abs/2006Natur.440.1137S} {440, 1137}

\bibitem[\protect\citeauthoryear{{Tollet}, {Cattaneo}, {Mamon}, {Moutard}  \&
  {van den Bosch}}{{Tollet} et~al.}{2017}]{Tollet17}
{Tollet} E.,  {Cattaneo} A.,  {Mamon} G.~A.,  {Moutard} T.,   {van den Bosch}
  F.~C.,  2017, \mn@doi [\mnras] {10.1093/mnras/stx1840}, \href
  {https://ui.adsabs.harvard.edu/abs/2017MNRAS.471.4170T} {471, 4170}

\bibitem[\protect\citeauthoryear{{Tulin} \& {Yu}}{{Tulin} \&
  {Yu}}{2018}]{Tulin18}
{Tulin} S.,  {Yu} H.-B.,  2018, \mn@doi [\physrep]
  {10.1016/j.physrep.2017.11.004}, \href
  {https://ui.adsabs.harvard.edu/abs/2018PhR...730....1T} {730, 1}

\bibitem[\protect\citeauthoryear{{Tulin}, {Yu}  \& {Zurek}}{{Tulin}
  et~al.}{2013}]{Tulin13}
{Tulin} S.,  {Yu} H.-B.,   {Zurek} K.~M.,  2013, \mn@doi [\prd]
  {10.1103/PhysRevD.87.115007}, \href
  {https://ui.adsabs.harvard.edu/abs/2013PhRvD..87k5007T} {87, 115007}

\bibitem[\protect\citeauthoryear{{Valli} \& {Yu}}{{Valli} \&
  {Yu}}{2018}]{Valli18}
{Valli} M.,  {Yu} H.-B.,  2018, \mn@doi [Nature Astronomy]
  {10.1038/s41550-018-0560-7}, \href
  {https://ui.adsabs.harvard.edu/abs/2018NatAs...2..907V} {2, 907}

\bibitem[\protect\citeauthoryear{{Vogelsberger}, {Zavala}  \&
  {Loeb}}{{Vogelsberger} et~al.}{2012}]{Vogelsberger12}
{Vogelsberger} M.,  {Zavala} J.,   {Loeb} A.,  2012, \mn@doi [\mnras]
  {10.1111/j.1365-2966.2012.21182.x}, \href
  {https://ui.adsabs.harvard.edu/abs/2012MNRAS.423.3740V} {423, 3740}

\bibitem[\protect\citeauthoryear{{Vogelsberger}, {Zavala}, {Cyr-Racine},
  {Pfrommer}, {Bringmann}  \& {Sigurdson}}{{Vogelsberger}
  et~al.}{2016}]{Vogelsberger16}
{Vogelsberger} M.,  {Zavala} J.,  {Cyr-Racine} F.-Y.,  {Pfrommer} C.,
  {Bringmann} T.,   {Sigurdson} K.,  2016, \mn@doi [\mnras]
  {10.1093/mnras/stw1076}, \href
  {https://ui.adsabs.harvard.edu/abs/2016MNRAS.460.1399V} {460, 1399}

\bibitem[\protect\citeauthoryear{{Vogelsberger}, {Zavala}, {Schutz}  \&
  {Slatyer}}{{Vogelsberger} et~al.}{2019}]{Vogelsberger19}
{Vogelsberger} M.,  {Zavala} J.,  {Schutz} K.,   {Slatyer} T.~R.,  2019,
  \mn@doi [\mnras] {10.1093/mnras/stz340}, \href
  {https://ui.adsabs.harvard.edu/abs/2019MNRAS.484.5437V} {484, 5437}

\bibitem[\protect\citeauthoryear{{Wetzel}, {Hopkins}, {Kim},
  {Faucher-Gigu{\`e}re}, {Kere{\v{s}}}  \& {Quataert}}{{Wetzel}
  et~al.}{2016}]{Wetzel16}
{Wetzel} A.~R.,  {Hopkins} P.~F.,  {Kim} J.-h.,  {Faucher-Gigu{\`e}re} C.-A.,
  {Kere{\v{s}}} D.,   {Quataert} E.,  2016, \mn@doi [\apjl]
  {10.3847/2041-8205/827/2/L23}, \href
  {https://ui.adsabs.harvard.edu/abs/2016ApJ...827L..23W} {827, L23}

\bibitem[\protect\citeauthoryear{{Wittman}, {Golovich}  \& {Dawson}}{{Wittman}
  et~al.}{2018}]{Wittman18}
{Wittman} D.,  {Golovich} N.,   {Dawson} W.~A.,  2018, \mn@doi [\apj]
  {10.3847/1538-4357/aaee77}, \href
  {https://ui.adsabs.harvard.edu/abs/2018ApJ...869..104W} {869, 104}

\bibitem[\protect\citeauthoryear{{Zavala}, {Vogelsberger}  \&
  {Walker}}{{Zavala} et~al.}{2013}]{Zavala13}
{Zavala} J.,  {Vogelsberger} M.,   {Walker} M.~G.,  2013, \mn@doi [\mnras]
  {10.1093/mnrasl/sls053}, \href
  {https://ui.adsabs.harvard.edu/abs/2013MNRAS.431L..20Z} {431, L20}

\bibitem[\protect\citeauthoryear{{van den Bosch}}{{van den
  Bosch}}{2017}]{vandenBosch17}
{van den Bosch} F.~C.,  2017, \mn@doi [\mnras] {10.1093/mnras/stx520}, \href
  {https://ui.adsabs.harvard.edu/abs/2017MNRAS.468..885V} {468, 885}

\bibitem[\protect\citeauthoryear{{van den Bosch}, {Ogiya}, {Hahn}  \&
  {Burkert}}{{van den Bosch} et~al.}{2018}]{vandenBosch18}
{van den Bosch} F.~C.,  {Ogiya} G.,  {Hahn} O.,   {Burkert} A.,  2018, \mn@doi
  [\mnras] {10.1093/mnras/stx2956}, \href
  {https://ui.adsabs.harvard.edu/abs/2018MNRAS.474.3043V} {474, 3043}

\makeatother
\end{thebibliography}
\bibliographystyle{mnras}

\appendix

\section{Gravothermal collapse model: Further details}\label{Gravothermal_model_appendix}

In this Appendix we provide additional details on the numerical implementation for solving the gravothermal collapse model introduced in Section~\ref{Sec_Gravothermal_collapse}. 

We begin by defining a new set of dimensionless variables based on the mass and length scales, $M_{0}=4\pi r_{s}^{3}\rho_{s}$ and $R_{0}=r_{s}$, where $\rho_{s}$ and $r_{s}$ are the scale density and radius, respectively, of the initial NFW profile. These quantities lead to consistent normalization scales for the other variables: $\rho_{0}=\rho{s}$, $v_{0}^{2}=GM_{0}/R_{0}$ (with G gravitational constant), $\sigma_{0}=4\pi R_{0}^{2}M_{0}^{-1}$, $L_{0}=G M_{0}^{2}/(t_{0}R_{0})$ and $t_{0}^{-1}=a\sigma_{m}v_{0}\rho_{0}$. 

Nondimensionless variables result $\tilde{r}=r/R_{0}$, $\tilde{m}=m/M_{0}$, $\tilde{v}=v/v_{0}$, $\tilde{L}=L/L_{0}$, $\tilde{\sigma}_{m}=\sigma_{m}/\sigma_{0}$ and $\tilde{t}=t/t_{0}$, which are then used to rewrite eqs.~(\ref{mass_conservation_eq}-\ref{flux_eq_2}) in a dimensionless form

\begin{eqnarray}\label{mass_conservation_eq_2}
\frac{\partial \tilde{m}}{\partial \tilde{r}} &=& \tilde{r}^{2}\tilde{\rho},\\\label{hydro_equi_eq_2}
\frac{\partial (\tilde{\rho}\tilde{v}^{2})}{\partial \tilde{r}} &=& -\frac{\tilde{m}\tilde{\rho}}{\tilde{r}},\\\label{flux_eq_3}
\tilde{L} &=& -\frac{3}{2}\tilde{r}^{2}\tilde{v}\left(\frac{a}{b}\tilde{\sigma}_{m}^{2}+\frac{1}{C\tilde{\rho}\tilde{v}^{2}}\right)^{-1}\frac{\partial \tilde{v}^{2}}{\partial \tilde{r}},\\\label{eq_4}
\frac{\partial \tilde{L}}{\partial \tilde{r}} &=& -\tilde{r}^{2}\tilde{\rho} \tilde{v}^{2}\left(\frac{\partial}{\partial \tilde{t}}\right)_{m}\log\left(\frac{\tilde{v}^{3}}{\tilde{\rho}}\right).
\end{eqnarray}

The initial density profile is used to calculate $\tilde{m}$ and $\tilde{v}$ through eqs.~(\ref{mass_conservation_eq_2}-\ref{hydro_equi_eq_2}). Those, along with the cross section, are used to calculate $\tilde{L}$ from eq.~(\ref{flux_eq_3}). We allow a small passage of time $\Delta\tilde{t}$ (given by eq.~\ref{timestep}) and compute the new density $\tilde{\rho}$ that solves eq.~(\ref{eq_4}). After this we go back to solving eqs~(\ref{mass_conservation_eq_2}-\ref{flux_eq_3}).

\section{Impact of initial conditions: changing $c_{200}$}\label{Comparison_c200_appendix}

\begin{figure} 
\begin{center}
	\includegraphics[angle=0,width=0.48\textwidth]{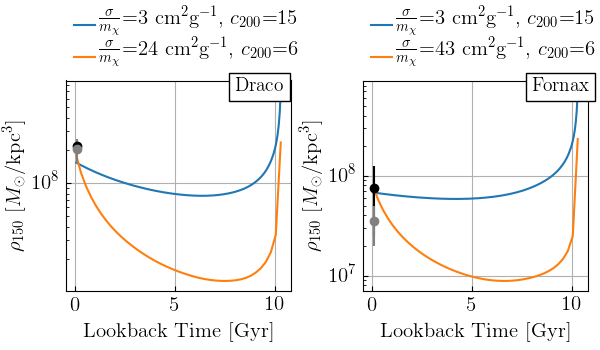}
	\caption{Draco's (left) and Fornax's (right) DM density at 150 pc, $\rho_{150}$, as a function of lookback time. In the left panel the coloured lines correspond to the subhalo model initialised with a cross section of $\sigma/m_{\chi}=3$ cm$^{2}$g$^{-1}$ and concentration parameter of $c_{200,\rm{init}}=15$ (blue line) and $\sigma/m_{\chi}=24$ cm$^{2}$g$^{-1}$ and $c_{200,\rm{init}}=6.3$ (orange line). Similarly, in the right panel the model was initialised with a cross section of $\sigma/m_{\chi}=3$ cm$^{2}$g$^{-1}$ and concentration parameter of $c_{200,\rm{init}}=15$ (blue line) and $\sigma/m_{\chi}=43$ cm$^{2}$g$^{-1}$ and $c_{200,\rm{init}}=6.3$ (orange line). The figure shows the large impact of the concentration parameter that initializes the NFW profile. If higher values of $c_{200}$ were assumed as starting point, lower values of $\sigma/m_{\chi}$ would be needed to reproduce $\rho_{150}$ (shown in symbols), as reported by Kaplinghat et al. (2019), who assumed an isothermal cored profile (grey symbol) as well as NFW (black symbol).}
	\label{Comparison_c200}
\end{center}
\end{figure}

The core collapse time-scale is shorter for low-concentration systems. In Section~\ref{Sec_impact_ICs_2} we discussed how changing the initial concentration parameters in our model impacts our results. Here we show an specific example. We compare the evolution of the central DM density, $\rho_{150}$, from Draco and Fornax, models that were run with different initial concentrations and cross sections. 

Fig.~\ref{Comparison_c200} shows $\rho_{150}$ as a function of lookback time for Draco (left panel) and Fornax (right panel). The lines in the panels correspond to the models initialised with a cross section of $\sigma/m_{\chi}=3$ cm$^{2}$g$^{-1}$ and concentration parameter of $c_{200,\rm{init}}=15$ (blue lines), and $\sigma/m_{\chi}=24$ cm$^{2}$g$^{-1}$ (43 cm$^{2}$g$^{-1}$ for Fornax) and $c_{200,\rm{init}}=6.3$ (orange lines). It can be seen that the concentration parameter largely impacts the evolution of $\rho_{150}$. We conclude that if higher values of $c_{200,\rm{init}}$ were assumed as a starting point, lower values of $\sigma/m_{\chi}$ would be needed to reproduce $\rho_{150}$ (shown in symbols taken from Kaplinghat et al. 2019).

\section{Impact of initial conditions: changing $M_{\rm{init}}$}\label{Comparison_Minit_appendix}

Section~\ref{Sec_central_density_evolution} concludes that all MW dSphs need to be in gravothermal core collapse in order to explain the observational data. We commented that this result, however, strongly depends on the initial virial mass of the systems, $M_{200,\rm{init}}$, which is not chosen at random, it is tuned so that the systems, in their final state, have a virial mass that reproduces the observational estimations.

\begin{figure} 
	\includegraphics[angle=0,width=0.48\textwidth]{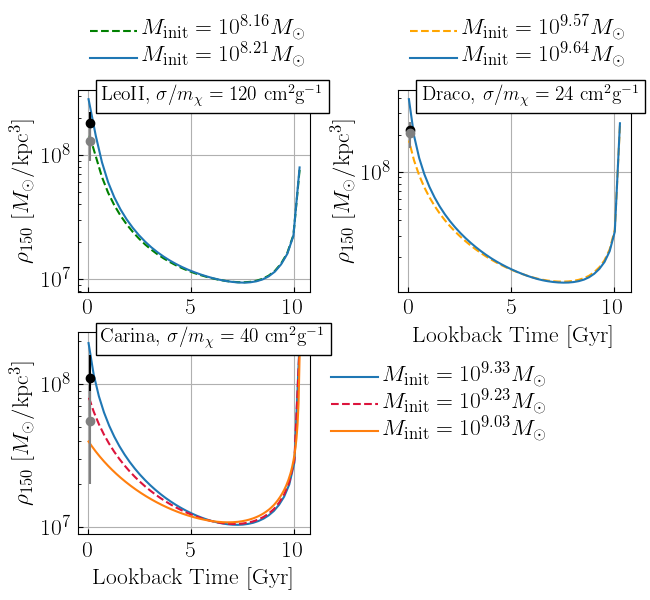}
	\caption{DM density at 150 pc, $\rho_{150}$, as a function of lookback time for the models of LeoII (top-left panel), Draco (top-right panel) and Carina (bottom-left panel). The coloured lines correspond to the subhaloes initialised with the same cross section ($\sigma/m_{\chi}=120$ cm$^{2}$g$^{-1}$ for LeoII, $24$ cm$^{2}$g$^{-1}$ for Draco and $40$ cm$^{2}$g$^{-1}$ for Carina), but different initial virial masses. The symbols show the values of $\rho_{150}$ (and $M_{200}$) taken from Kaplinghat et al. (2019), who assumed an isothermal cored profile (grey symbol) as well as NFW (black symbol). The figure shows how the initial mass of the models impact on the final DM density profile.}
	\label{Comparison_Initial_Mass}
\end{figure}

In this section we show that changing the initial virial masses of systems such as LeoII, Draco and Carina in $20\%$, changes the final central DM densities of these subhaloes in up to $50\%$. Fig.~\ref{Comparison_Initial_Mass} shows the DM density at 150 pc, $\rho_{150}$, as a function of lookback time for the models of LeoII (top-left panel), Draco (top-right panel) and Carina (bottom-left panel). The coloured lines correspond to the subhaloes initialised with the same cross section ($\sigma/m_{\chi}=120$ cm$^{2}$g$^{-1}$ for LeoII, $24$ cm$^{2}$g$^{-1}$ for Draco and $40$ cm$^{2}$g$^{-1}$ for Carina), but different initial virial masses. 

The top panels show that the models of LeoII and Draco lower their final DM densities from $3\times 10^{8}\Msun$ kpc$^{-3}$ to $1.5\times 10^{8}\Msun$ kpc$^{-3}$, and from $4\times 10^{8}\Msun$ kpc$^{-3}$ to $2\times 10^{8}\Msun$ kpc$^{-3}$, respectively. The bottom panels shows that for Carina, an initial virial mass of $10^{9.3}\Msun$ leads to a final DM density of $2\times 10^{8}\Msun$ kpc$^{-3}$, lowering the initial mass by $20\%$ results in a DM density of $9\times 10^{7}\Msun$ kpc$^{-3}$, and further lowering the initial mass by $50\%$ results in a DM density of $4\times 10^{7}\Msun$ kpc$^{-3}$.

\section{Impact of Milky Way mass}\label{Comparison_MWMass_appendix}

The mass of the Milky Way (MW) can impact the results presented in Section~\ref{Results} by altering the orbital evolution of the dwarfs (shown in Fig.~\ref{OrbitsDwarfGalaxies}), as well as the rate of mass loss by increasing/decreasing the tidal radius above which mass is tidally stripped. In this Section we investigate how changing the MW mass from $10^{12}\Msun$ (default value adopted throughout the work) to $0.8\times 10^{12}\Msun$ or $1.6\times 10^{12}\Msun$, changes the final estimates of the DM density of dwarf subhaloes undergoing gravothermal collapse.

\begin{figure} 
\begin{center}
	\includegraphics[angle=0,width=0.35\textwidth]{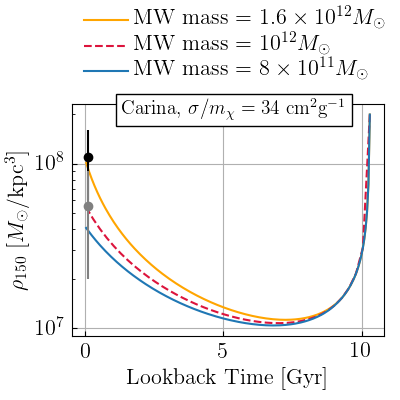}
	\caption{Carina's DM density at 150 pc, $\rho_{150}$, as a function of lookback time. The coloured lines correspond to the subhalo model initialised with a cross section of $\sigma/m_{\chi}=34$ cm$^{2}$g$^{-1}$, but different values for MW mass. The symbols show the values of $\rho_{150}$ (and $M_{200}$) taken from Kaplinghat et al. (2019), who assumed an isothermal cored profile (grey symbol) as well as NFW (black symbol). The figure shows that assuming a MW mass of $1.6\times 10^{12}\Msun$ ($0.8\times 10^{12}\Msun$) requires a factor of $\approx 1.2$ lower (higher) $\sigma/m_{\chi}$ to reproduce the observed central DM densities.}
	\label{Comparison_MW_mass_fig}
	\end{center}
\end{figure}

Fig.~\ref{Comparison_MW_mass_fig} shows Carina's DM density at 150 pc, $\rho_{150}$, as a function of lookback time. The various subhalo models shown in the figure were all initialised with a cross section of $\sigma/m_{\chi}=34$ cm$^{2}$g$^{-1}$ and a MW mass of: $0.8\times 10^{12}\Msun$ (shown in blue solid line), $10^{12}\Msun$ (red dashed line) and $1.6\times 10^{12}\Msun$ (yellow solid line). The figure shows that a high MW mass, induces a higher rate of mass loss from gravitational tidal stripping, and therefore further accelerates the subhaloes' gravothermal collapse, resulting in subhaloes with higher central DM densities. On the contrary, a lower MW mass produces subhaloes with lower central DM densities. 

For Carina, but this also applies to the other dwarf models, increasing (decreasing) the MW mass in a factor of 1.6 (1.2), increases (decreases) the central DM density in a factor of 2 (1.2). This implies that assuming a MW mass of $1.6\times 10^{12}\Msun$, requires a factor of up to $\approx 1.2$ lower cross sections to reproduce the observed central DM densities, or alternatively assuming a MW mass of $0.8\times 10^{12}\Msun$, requires a factor of $\approx 1.2$ larger cross sections.

\section{Impact of Truncation time}\label{Impact_trunctation_time}

The truncation time, $t_{\rm{trunc}}$, is a free parameter that regulates the frequency over which the subhaloes' density profile is truncated. In Section~\ref{Sec_Tidal_Stripping} we showed that the largest changes of the density profile occurs in the outskirt of the halo, beyond the virial radius, such change in the profile is, nevertheless, important because it lowers the pressure of the extended halo, it gets therefore colder as it readjusts to hydrostatic equilibrium. Section~\ref{Sec_central_density_evolution} shows that this favours the conditions for gravothermal core collapse.

\begin{figure} 
\begin{center}
	\includegraphics[angle=0,width=0.35\textwidth]{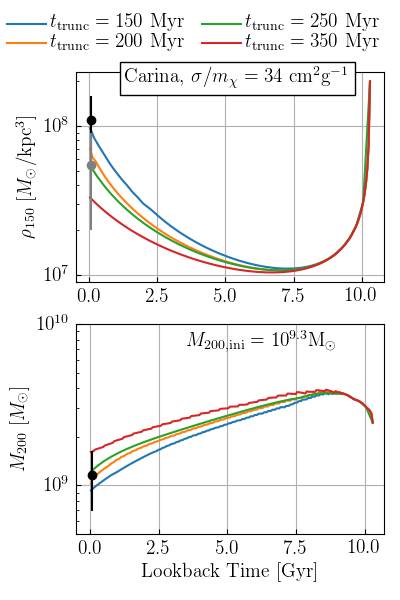}
	\caption{Carina's DM density at 150 pc, $\rho_{150}$ (top panel), and virial mass, $M_{200}$ (bottom panel), as a function of lookback time. The coloured lines correspond to the subhalo model initialised with the same cross section (of $\sigma/m_{\chi}=34$ cm$^{2}$g$^{-1}$) and initial mass, but different values for the truncation time, $t_{\rm{trunc}}$, that determines the frequency over which subhaloes' density profile is truncated due to mass loss. The symbols show the values of $\rho_{150}$ (and $M_{200}$) taken from Kaplinghat et al. (2019), who assumed an isothermal cored profile (grey symbol) as well as NFW (black symbol). The figure shows that a more frequent truncation, accelerates gravothermal core collapse and subhaloes reach higher $\rho_{150}$ at present time. Conversely, a less frequent truncation of the density profile, decelerates gravothermal core collapse and subhaloes reach lower $\rho_{150}$.}
	\label{Comparison_Truncation_times}
	\end{center}
\end{figure}

In this section we investigate how changing $t_{\rm{trunc}}$ from 250 Myr (default value) to 350 or 150 Myr, alters the final central DM density of subhaloes. Fig.~\ref{Comparison_Truncation_times} shows the Carina model initialised with a cross section of $\sigma/m_{\chi}=34$ cm$^{2}$g$^{-1}$. The top panel shows the DM density at 150 pc, $\rho_{150}$, as a function of lookback time, whereas the bottom panel shows the evolution of the virial mass, $M_{200}$. The coloured lines correspond to the subhalo model initialised with the same cross section and initial mass, but different values for the truncation time as indicated in the legends. It can be seen from the figure that a more frequent truncation accelerates gravothermal core collapse and subhaloes reach higher $\rho_{150}$ at present time. Conversely, a less frequent truncation of the density profile, decelerates gravothermal core collapse and subhaloes reach lower $\rho_{150}$.

For the specific cross section of $\sigma/m_{\chi}=34$ cm$^{2}$g$^{-1}$, the models with different $t_{\rm{trunc}}$ reach $\rho_{150}$ and $M_{200}$ at present time in agreement with the observational estimations (within the uncertainties). Lowering $t_{\rm{trunc}}$ to 150 Myr, yields $50\%$ higher central DM densities, while increasing $t_{\rm{trunc}}$ to 350 Myr, yields $30\%$ lower central DM densities. This implies that changing $t_{\rm{trunc}}$ from 250 to 150 Myr (or 350 Myr), increases (decreases) the cross sections range that reproduces the observed central DM densities in up to a factor of 1.12 (1.13). Therefore the main effect of $t_{\rm{trunc}}$ in our results is to increase or decrease the normalization of the cross section-velocity relation, but it does not alters the shape of the relation.

\end{document}